\documentclass[aps,twocolumn,
groupedaddress,longbibliography,prl]{revtex4-2}

\usepackage{amsmath,amssymb}
\usepackage{graphicx}
\usepackage{multirow}
\usepackage{bm}
\usepackage{mathtools}
\usepackage{dsfont}
\usepackage{amsfonts}
\usepackage{xcolor}
\usepackage[normalem]{ulem}
\usepackage{url}
\usepackage{wasysym}
\usepackage{bbm}

\usepackage[colorlinks=true,linkcolor=magenta,citecolor=magenta,hypertexnames=false]{hyperref}
\newcommand{\nocontentsline}[3]{}
\newcommand{\tocless}[2]{\bgroup\let\addcontentsline=\nocontentsline#1{#2}\egroup}

\def\ba#1\ea{\begin{align}#1\end{align}}
\def\bg#1\eg{\begin{gather}#1\end{gather}}
\def\bpm{\begin{pmatrix}}
\def\epm{\end{pmatrix}}

\newcommand{\ket}[1]{|#1\rangle}
\newcommand{\bra}[1]{\langle#1|}

\newcommand{\magenta}[1]{\textcolor{magenta}{#1}}

\newcommand{\ourtitle}{
Strain Response as a Probe of Spinons in Quantum Spin Liquids}

\allowdisplaybreaks

\begin{document}
\title{\textbf{\ourtitle}}


 \author{
Penghao Zhu$^{1}$, 
Archisman Panigrahi$^{2}$, Leonid Levitov$^{2}$, and Nandini Trivedi$^{1}$
}
%

\affiliation{$^1$Department of Physics, The Ohio State University, Columbus, OH 43210, USA\\
$^2$Department of Physics, Massachusetts Institute of Technology, Cambridge, MA 02139, USA}


\begin{abstract}
Quantum spin liquids (QSLs) host emergent, fractionalized fermionic excitations that are charge-neutral. Identifying clear experimental signatures of these excitations remains a central challenge in the field of strongly correlated systems, as they do not couple to conventional electromagnetic probes. Here, we propose lattice strain as a powerful and tunable probe: Mechanical deformation of the lattice generates large pseudomagnetic fields, inducing pseudo–Landau levels that serve as distinctive spectroscopic signatures of these excitations. Using the Kitaev model on the honeycomb lattice, we show that distinct QSL phases exhibit strikingly different strain responses. The semimetallic Kitaev spin liquid and the gapped chiral spin liquid display pronounced Landau quantization and a diamagnetic-like response to strain, whereas the Majorana metal phase shows a paramagnetic-like response without forming Landau levels. These contrasting behaviors provide a direct route to experimentally identifying and distinguishing QSL phases hosting fractionalized excitations. We further outline how local resonant ultrasound spectroscopy can detect the strain-induced resonances associated with these responses, offering a practical pathway towards identifying fractionalized excitations in candidate materials.
\end{abstract}

\maketitle

\let\oldaddcontentsline\addcontentsline
\renewcommand{\addcontentsline}[3]{}

The most fascinating aspect of quantum spin liquids (QSLs) is that they harbor fractionalized neutral excitations(spinons), which may be complex or Majorana. However, identifying definitive observable consequences of fractionalized degrees of freedom remains a central challenge in this field. The observation of broad features in inelastic neutron and Raman scattering spectra~\cite{han2012,banerjee2016,wulferding2010,gupta2016} have provided supportive evidence of the fractionalization of the usual magnon modes in a quantum spin liquid, and there are some hints of observation of quantized thermal Hall conductance~\cite{kasahara2018} and quantum oscillations in magnetothermal transport~\cite{czajka2021} in $\alpha -{\rm RuCl_3}$ as well as Friedel-like oscillations observed via scanning tunneling microscopy (STM)~\cite{Kohsaka2024}, these results are still under debate, especially their implications toward fractionalization.

\begin{figure}[t!]
\centering
\includegraphics[width=1\columnwidth]{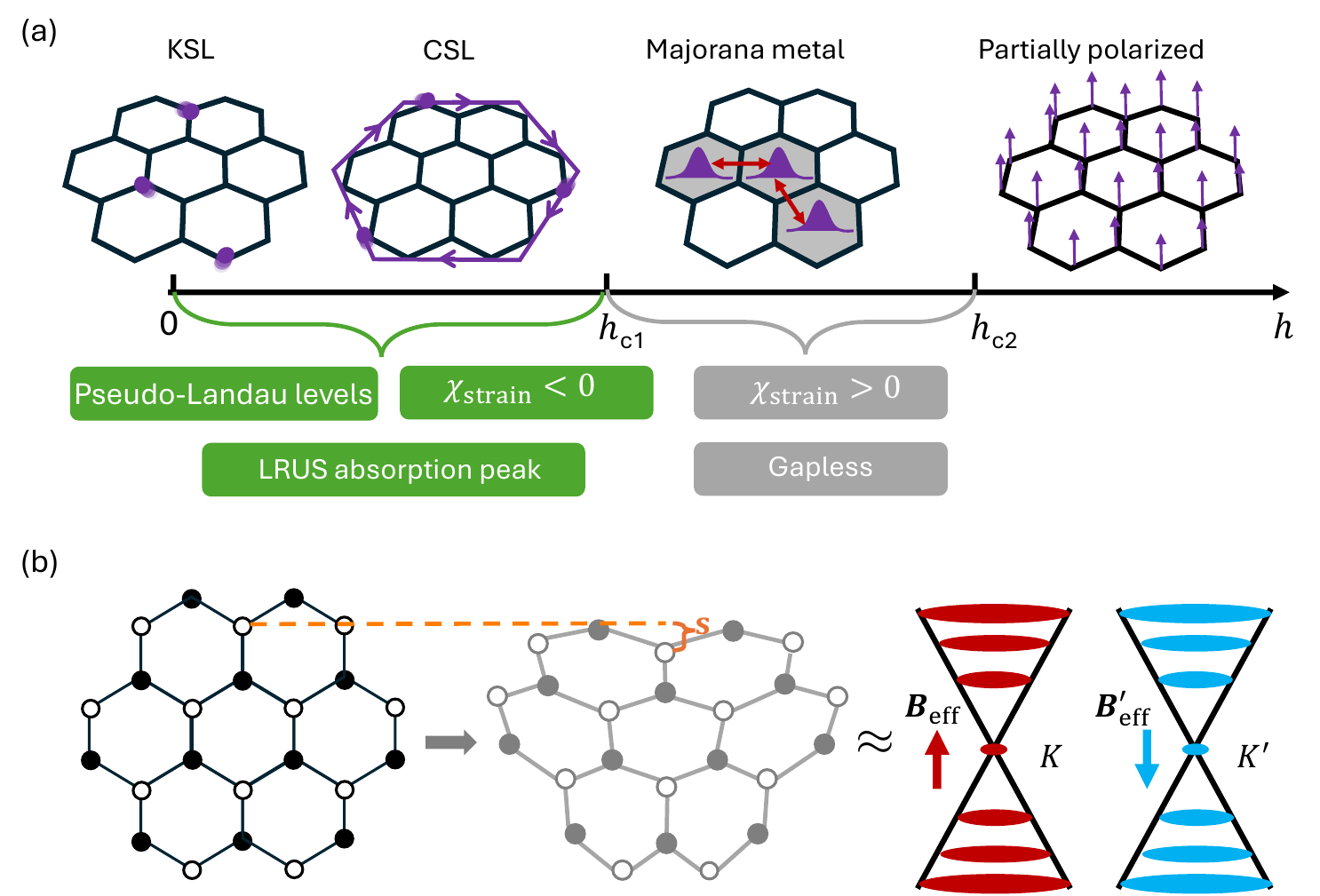}
\caption{(a) Phase diagram of the antiferromagnetic Kitaev honeycomb model under an external magnetic field [see Eq.\eqref{eq:H}] and the strain responses of the three quantum spin liquid phase. KSL denotes the Kitaev spin liquid at the exactly solvable limit $h=0$, while CSL denotes the chiral spin liquid phase at small $h$. LRUS denotes local resonant ultrasound spectroscopy. 
(b) Strain configuration described by Eq.\eqref{eq:strain} and strain induced pseudomagnetic field. Black dots and circles represent the unstrained lattice, while gray ones show the deformed positions after applying strain. The strain induces opposite pseudomagnetic field at $K$ and $K'$.}
\label{fig1}
\end{figure}

One major obstacle to detecting spinons is that these quasiparticles are charge neutral and thus do not couple directly to external electromagnetic fields, eliminating a broad and powerful class of conventional probes. 
For charged fermions, coupling to an external magnetic field is a prime tool to probe phase-coherent dynamics—from quantum oscillations and Landau quantization to integer and fractional quantum Hall effects. Identifying analogous synthetic gauge fields that allow similar probes for charge-neutral systems is a question of considerable interest, one that has so far received limited attention in the literature.

In this article, we turn to external strain fields that can couple to neutral excitations and propose strain-induced synthetic gauge fields as a vehicle to probe correlated states of QSLs that host charge-neutral fermionic excitations. The coupling between lattice deformation and fermions is usually given by the deformation potential, which acts as a scalar potential. However, when the relevant fermionic excitations are located away from the Brillouin-zone center, an additional type of coupling emerges—one that enters the Hamiltonian as minimal coupling $H(\mathbf{p}-\mathbf{A}_{\text{eff}})$ where $\mathbf{A}_{\text{eff}}$ is a vector potential linear in the strain field. A uniform, coordinate-independent $\mathbf{A}_{\text{eff}}$ simply shifts the momentum and has no effect on quasiparticle dynamics. In contrast, a spatially varying strain produces a nonuniform vector potential $\mathbf{A}_{\text{eff}}$, which acts as a pseudomagnetic field experienced by the fermions and has been a fruitful avenue of investigation in recent years. To the best of our knowledge, it was first discussed by Iordanskii and Koshelev \cite{IordanskiiKoshelev1985JETPLett}, who predicted that dislocations in silicon crystals act as pseudo-Aharonov-Bohm solenoids, carrying a geometric magnetic flux determined by the Burgers vector of the dislocation. The idea was later developed for graphene under strain, where elastic deformation can produce pseudomagnetic fields that couple to Dirac electrons \cite{Abanin2007, guinea2010, levy2010, Berman2022}. It was invoked to predict order-from-disorder effects in quantum Hall ferromagnets in rippled graphene\cite{Abanin2007}. Subsequently, in a work that gained much publicity, these fields were predicted to give rise to pseudo-Landau levels in the quasiparticle spectrum and, for sufficiently strong strain, can reach effective field strengths on the order of hundreds of tesla \cite{guinea2010,levy2010}. More recently, pseudomagnetic fields have been exploited as a tunable parameter to modify the spectra of particle–hole excitations (such as Wannier excitons) in strained graphene \cite{Berman2022}, providing a novel route to engineer electronic and optical responses in two-dimensional materials.

We extend these ideas to QSLs and investigate the strain responses of the Kitaev model on the honeycomb lattice under different external magnetic fields [see Fig.~\ref{fig1}(a)],
where the coupling of spinons to strain takes the similar geometric form as in Dirac materials. We show that the strain response of the Kitaev honeycomb model exhibits qualitatively distinct features under different external magnetic fields, reflecting the properties of fractionalized excitations across the various QSL phases. We demonstrate that these strain responses affect the experimentally measured elastic tensors, which can be probed using resonant ultrasound spectroscopy (RUS)~\cite{schwarz2000,dukhin2002,balakirev2019,hauspurg2024,Theuss2024_2}, and can serve as a signature of fractionalized excitations in candidate spin-liquid materials. Specifically, in the low-field phases, pseudo-Landau levels emerge as expected and result in a negative static strain susceptibility, whereas in the intermediate-field gapless phase—despite the presence of a Fermi surface—no pseudo-Landau levels appear, and the static strain susceptibility is positive [see Fig.~\ref{fig1}(a)]. The pseudo-Landau level spacing can be detected through ultrasound attenuation capturing the absorption of the ultrasound wave, and the static strain susceptibility, which is closely related to the elastic tensor, can be measured with RUS.

\noindent \magenta{\it Phases of the Kitaev Honeycomb model in an external field:|} In Kitaev’s honeycomb model, the spin $1/2$ local moments on lattice sites interact via Ising interactions involving non-commuting Pauli operators on the three types of bonds. In this case the interaction-driven frustration leads to the fractionalization of the spin operator into two new charge-neutral quasiparticles that act as itinerant matter Majoranas and emergent gauge fields, respectively~\cite{kitaev2006anyons}. 

Since an external magnetic field is often required to stabilize quantum spin liquid (QSL) phases in candidate Kitaev materials, we focus on the Kitaev model on a honeycomb lattice,  with a magnetic field applied normal to the honeycomb plane:
\begin{equation}
\label{eq:H}
    H = \sum_{\langle ij\rangle_\alpha} J_{\alpha}{\sigma^\alpha_i \sigma^\alpha_j} - \sum_{i,\alpha} h \sigma_i^\alpha ,~~\alpha = x,y,z
\end{equation}
where $\langle ij\rangle_\alpha$ denotes  nearest-neighbor sites on an $\alpha$-type bond. In the isotropic limit with antiferromagnetic Kitaev interactions, i.e., $J_x=J_y=J_z>0$, model in Eq.~\eqref{eq:H} exhibits three QSL phases [see Fig.~\ref{fig1}(a)]: When $h=0$ at the exactly solvable limit, we have a gapless QSL phase featured by the low-energy Dirac modes, which is referred to as Kitaev spin liquid (KSL). At small $h$, perturbations from the Zeeman term open a topological gap, leading to a gapped chiral spin liquid (CSL) phase featured by the gapless edge chiral Majoranas. At intermediate field strengths, there is a gapless QSL phase which is recently identified as a Majorana metal phase by two of the authors~\cite{wang2025,zhu2025,feng2025}. The Majorana metal phase emerges due to the proliferation of $\mathbb{Z}_2$ fluxes in the ground state. These fluxes trap Majorana zero modes, which then hybridize and form a gapless band. In the following, we explore the strain responses of these three QSL phases.

\begin{figure}[h]
\centering
\includegraphics[width=0.6\columnwidth]{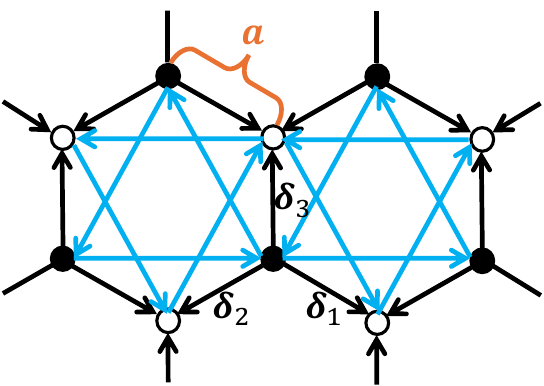}
\caption{Majorana hoppings on the honeycomb lattice depicted in Eq.~\eqref{eq:tbH}.}
\label{fighopping}
\end{figure}

\noindent \magenta{\it Strain field in Kitaev's model and pseudo-Landau levels.|}
We begin by describing how strain modifies the model in Eq.~\eqref{eq:H}. When lattice sites are displaced by a vector field $\mathbf{s}$ [see Fig.~\ref{fig1}(b)], the strain tensor is defined as $\epsilon_{ij}=\partial_{i}s_{j}$. 
When a chemical bond is distorted by strain, the hopping amplitude $J_0$ between neighboring orbitals typically changes because the orbital overlap depends critically on the inter-atomic distance. A widely used parametrization is \cite{Ribeiro_strained_2009, Pearce_tight-binding_2016},
\begin{equation} 
\label{eq:strainedinteraction} 
J_{\alpha} = J_{0}\exp\left[-\beta\left(\frac{|(I + \epsilon)\boldsymbol{\delta}_{\alpha}|}{a} - 1\right)\right], \end{equation} 
where the vector $\boldsymbol{\delta}_{\alpha}$ connects nearest-neighbor sites on an $\alpha$-type bond and has equilibrium length $a$ [see Fig.~\ref{fighopping}], $|(I+\epsilon) \boldsymbol{\delta}_{\alpha}|$ is the strained length, and $\beta$ is a dimensionless decay parameter (a Gr\"{u}neisen-like constant). For small strains one often linearizes this as
\begin{equation} 
J_{\alpha} = J_{0}\left[1-\beta\left(\frac{|(I + \epsilon)\boldsymbol{\delta}_{\alpha}|-a}{a}\right)\right]. \end{equation} 
Eq.~\eqref{eq:strainedinteraction} reflects an exponential dependence of the exchange interaction on bond length, as is typical for superexchange. As discussed previously~\cite{kitaev2006anyons,zhu2025}, all the three quantum spin liquid phases can be efficiently captured by a Majorana hopping model
\begin{equation}
\label{eq:tbH}
H_{M}=\frac{1}{2}\sum_{j,k}it_{jk}c_jc_k,
\end{equation}
where $t_{jk}=\operatorname{sgn}(j,k)J_{\alpha_{jk}}u_{jk}$ for the nearest neighbor hoppings and $t_{jk}=\operatorname{sgn}(j,k)\lambda u_{jl}u_{lk}$ for next nearest neighbor hoppings, with $\lambda\sim h^{3}/(J_{\alpha_{jl}}J_{\alpha_{lk}})$ arising from the third order perturbation. $u_{jk}$ here is the expectation value of the $\mathbb{Z}_{2}$ vector gauge potential $\hat{u}_{jk}$ defined on the bond connecting sites $j$ and $k$. $\operatorname{sgn}(j,k)$ is $+1$ ($-1$) if the hopping is along (against) the direction of the arrow shown in Fig.~\ref{fighopping}. $\alpha_{jk}$ specifies the type of the $jk$ bond.  The strain enters the Majorana hopping model straightforwardly through $J_{\alpha}$.

In the absence of strain, the Majorana Hamiltonian in the flux-free sector (Eq.~\eqref{eq:tbH}) is known to exhibit low-energy Dirac modes at the high-symmetry points free $K$ and $K'$ in momentum space. The effective continuum Hamiltonian of these Dirac modes, up to a choice of basis, is given by
\begin{equation}
\label{eq:Diracmodes}
h_{\text{Dirac}}=3J_{0}a(k_x\sigma_x\pm k_{y}\sigma_y)\pm m\sigma_z
\end{equation}
where $\mathbf{k}=(k_x, k_y)$ the momentum relative to $K$ and/or $K'$, and $m=6\sqrt{3}\lambda$. The upper and lower signs correspond to the $K$ and $K'$ points, respectively. Since the physics at $K$ and $K'$ is similar, we focus on the Dirac mode around $K$ below. 

Consider the displacements, $\mathbf{s}(\mathbf{R})$, describing a biaxial strain applied to a flake of radius $L$ (the distance from the center to the boundary):
\begin{equation}
\label{eq:strain}
\mathbf{s}(\mathbf{R})=\frac{\mathcal{S}}{L}(2R_xR_y,R_x^2-R_y^2)
\end{equation}
where $\mathbf{R}$ is the coordinate of the lattice site defined with respect to the center of the center hexagon; and $\mathcal{S}$ a dimensionless parameter characterizing the strain strength. An illustration of this strain is shown in Fig.~\ref{fig1}(b).

\begin{figure}[h]
\centering
\includegraphics[width=1\columnwidth]{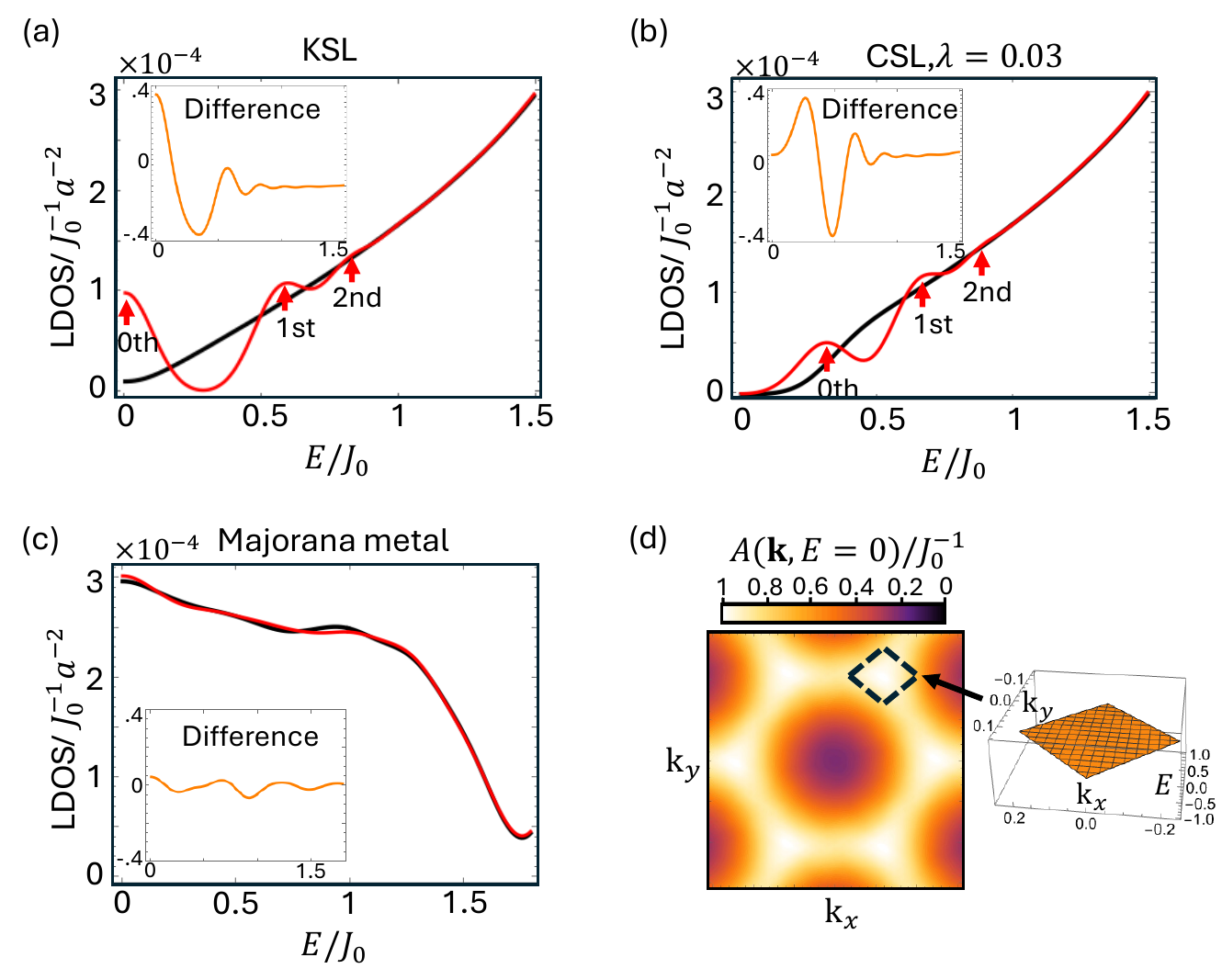}
\caption{Local density of states (LDOS) for region $\mathcal{R}$, defined as the central hexagon, in the (a) Kitaev spin liquid (KSL), (b) chiral spin liquid (CSL), and (c) Majorana metal phases. Red lines correspond to strained systems with $\beta \mathcal{S} = 0.4$, while black lines represent unstrained systems ($\beta \mathcal{S} = 0$). Insets show the LDOS difference between strained and unstrained systems. In the Majorana metal phase, we use $\lambda = 0.15 J_{0}$, disorder is introduced by flipping the sign of each bond with probability 0.2. The results are averaged over 100 samples. In all calculations we use $J_{0}=1$ and lattice with $60\times 60$ unit cells. (d) Momentum-resolved Majorana spectral function at zero energy ($E = 0$) and the dispersion near the $K$ point.}
\label{fig2}
\end{figure}

When this strain is present, it modifies the exchange couplings $J_{\alpha}$, which in turn alters the hopping amplitudes in the Majorana model. If we neglect changes to the Fourier phase factors, then to leading order, it is known that the strain field generates an effective vector gauge potential~\cite{suzuura2002,manes2007,guinea2010,masir2013}:
\begin{equation}
\label{eq:Aeff}
\begin{aligned}
\mathbf{A}_{\text{eff}}&=-\frac{\beta}{2 a}\frac{\hbar}{e}(\epsilon_{xx}-\epsilon_{yy}, -2\epsilon_{xy})
\\
&=\beta\mathcal{S}\frac{\hbar/e}{a L}(-2R_y, 2R_x),
\end{aligned}
\end{equation}
which corresponds to a uniform pseudomagnetic field $\mathbf{B}_{\text{eff}}=\nabla\times\mathbf{A}_{\text{eff}}=4\beta\mathcal{S}\frac{\hbar/e}{aL}\hat{z}$. The pseudomagnetic field reverses the sign between 
$K$ and $K'$ [see Fig.~\ref{fig1}(b)], which is consistent with the fact that strain field will not break the time-reversal symmetry. A detailed derivation of Eq.~\eqref{eq:Aeff} and the pseudomagnetic field can be found in Supplementary Material (SM). This pseudomagnetic field leads to pseudo-Landau levels in the KSL and CSL phases, where the $\mathbb{Z}_2$ flux remains gapped and is not significantly modified by the applied strain~\cite{yamada2023}. The spectrum is given by 
\begin{equation}
\varepsilon_{n,\pm,K} =\left\{\begin{array}{ll} m \ \ \text{when} \ n=0
\\
\pm \sqrt{m^2 +2 v_{\text{eff}}^2 \hbar e |\mathbf{B}_{\text{eff}}| n}  \ \ \text{otherwise}
\end{array}\right.
\end{equation}
with $v_{\text{eff}}=\frac{3J_{0}a}{\hbar}$. As shown in Fig.~\ref{fig2}(a)-(b), this spectrum is confirmed by our numerical calculations of the local density of states (LDOS). The local density of states in region $\mathcal{R}$ is defined as
\begin{equation}
\label{eq:LDOS}
\rho(E,\mathcal{R})=\sum_n\sum_{i\in \mathcal{R}}|\psi_{n,i}|^2\delta(E-\varepsilon_{n}),
\end{equation}
where $\psi_{n}$ and $\varepsilon_{n}>0$ are the wavefunction and energy of the $n$-th excited mode of the Majorana hopping model in Eq.~\eqref{eq:tbH}.

Given the gapless nature of the Majorana metal phase induced by emergent disordered fluxes in the intermediate field regime~\cite{zhu2025}, one might expect that a strain field would likewise generate pseudo-Landau levels, thereby enabling this novel phase to be identified and distinguished from the KSL and CSL phases through quantum oscillation measurements. However, as shown in our numerical results in Fig.\ref{fig2}(c), no pseudo-Landau levels emerge in this phase, which is effectively described by Eq.\eqref{eq:tbH} with sign randomness in $u_{ij}$. Although some oscillations are observed in the difference between the LDOS of strained and unstrained systems, they are relatively small compared to those seen in the KSL and CSL phases, and they lack universality, i.e., they are sensitive to the sample average process of the sign randomness in $u_{ij}$, of which the details can be found in SM. 

The absence of pseudo-Landau levels is straightforward to understand: extracting the dispersion near the gapless region by tracking the peaks in the spectral function, reveals that the low-energy modes are flat, as shown in Fig.~\ref{fig2}(d)~\footnote{Details about this calculation can be found in SM.}. This flatness is consistent with the logarithmic divergence of the density of states at low energies in the Majorana metal phase.  Consequently, strain does not couple to the orbital motion of the low energy modes and act as an effective gauge potential in the same way it does for linearly dispersing Dirac modes, and thus fails to generate pseudo-Landau quantization~\footnote{In general, when the Hamiltonian is not linear in momentum, the strain will generally not couple to the orbital motion like an effective vector gauge field, and can serve as a trivial perturbation as illustrated in SM.}. 

Meanwhile, quantum oscillation measurements are not straightforward to conduct, as neither the chemical potential of the itinerant Majoranas nor the strength of the pseudomagnetic field are easily tunable in a continuous way.

\noindent\magenta{\it Susceptibility to strain.|} Fortunately, despite the limitations of quantum oscillation measurements, the distinct spectral responses to the strain of the three phases manifest clearly in strain susceptibility and its temperature dependence. The strain susceptibility can serve as a useful probe for fractionalization and is defined through
\begin{equation}
\label{eq:strainsusceptibility}
\begin{aligned}
&M_{\text{eff}}(\mathcal{R},T)=-\partial\Omega_{G}(\mathcal{R},T)/\partial\mathcal{S},
\\
&\chi_{\text{strain}}(\mathcal{R},T)=\partial M_{\text{eff}}/\partial \mathcal{S},
\end{aligned}
\end{equation}
where $\Omega_G(\mathcal{R},T)$ denotes the local grand potential of the ground state in region $\mathcal{R}$, given by

\begin{equation}
\label{eq:grandp}
\begin{aligned}
&\Omega_{G}(\mathcal{R},T)=
\\
& \ E_{G}(\mathcal{R})+\sum_{n} -K_B T\log[1+e^{-\varepsilon_{n}/K_{B}T}]\sum_{i\in \mathcal{R}}|\psi_{n,i}|^2,
\end{aligned}
\end{equation}
with $E_{G}(\mathcal{R})$ being the local energy of the ground state, which is given by $E_{G}(\mathcal{R})=-1/2\sum_{n}\varepsilon_{n}\sum_{i\in \mathcal{R}}|\psi_{n,i}|^2$. The $1/2$ factor in $E_{G}(\mathcal{R})$ is due to the particle-hole symmetry, of which the details can be found in SM.

We refer to $M_{\text{eff}}$ and $\chi_{\text{strain}}$ as the pseudomagnetization and strain susceptibility, respectively, since in the KSL and CSL phases, the magnitude of strain field, $\mathcal{S}$, plays a role of a pseudomagnetic field. 
Beyond this analogy, the pseudomagnetization and strain susceptibility fundamentally represent the position-independent magnitude of the stress tensor $\Sigma^{ij}$ and the elasticity tensor $C^{ijkl}$ contributed by Majoranas:
\begin{equation}
\label{eq:elasticity}
\begin{aligned}
&\Sigma^{ij}=\frac{\partial\Omega_{G}}{\partial\epsilon_{ij}}=-M_{\text{eff}}\left(\frac{\partial\epsilon_{ij}}{\partial \mathcal{S}}\right)^{-1},
\\
&C^{ijkl}=\frac{\partial^{2}\Omega_{G}}{\partial\epsilon_{ij}\partial\epsilon_{kl}}=-\chi_{\text{strain}}\left(\frac{\partial\epsilon_{ij}}{\partial \mathcal{S}}\right)^{-1}\left(\frac{\partial\epsilon_{kl}}{\partial \mathcal{S}}\right)^{-1},
\end{aligned}
\end{equation}
where $\frac{\partial\epsilon_{ij}}{\partial\mathcal{S}}\propto \frac{R_{x,y}}{L}$ for the strain described in Eq.~\eqref{eq:strain}, serves as the position dependent part.

\begin{figure}[t]
\centering
\includegraphics[width=1\columnwidth]{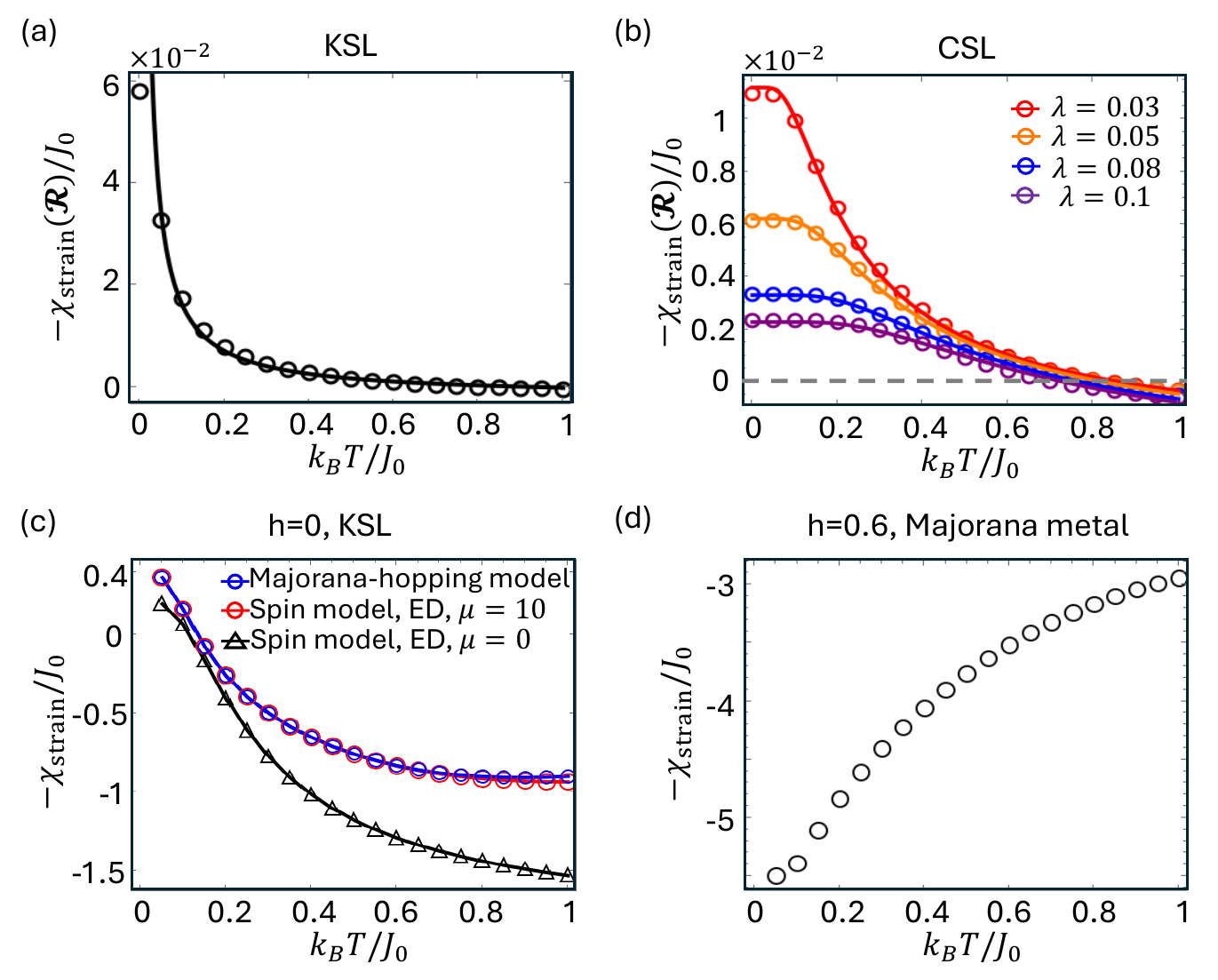}
\caption{Local strain susceptibility for region $\mathcal{R}$, defined as the central hexagon, in the (a) KSL and (b) CSL phases. Circles represent numerically data, solid line represents fittings of the data according to Eq.~\eqref{eq:suscsl}.  In both calculations we use $J_{0}=1$ and lattice with $60\times 60$ unit cells. (c) Strain susceptibility of a system with $4\times 2$ unit cells calculated from the spin model with and without the flux penalty and the Majorana-hopping model, respectively. (d) Strain susceptibility of a system with $4\times 2$ unit cells calculated from the spin model with $h=0.6$, which is in the Majorana metal phase.   }
\label{fig3}
\end{figure}

In the KSL and CSL phases, where the low-energy physics is essentially captured by Dirac modes coupled to effective vector gauge fields, the response to strain closely resembles the magnetic response in graphene. While the free energy of the fully filled Dirac cone is unbounded, the change in energy due to a small pseudomagnetic field remains finite, and the corresponding susceptibility remains finite. Specifically, at zero field we find $M_{\text{eff}}=0$ and a significant negative (``diamagnetic'') strain susceptibility $\chi_{\text{strain}}(\mathcal{R},T)$ at low temperature~\cite{gomez2011, ominato2013,vallejo2021}. We compute the strain-induced change in energy of the pseudoLandau levels, and obtain the following characteristic temperature dependence for the strain susceptibility (see SM for details):
\begin{equation}
\label{eq:suscsl}
 \chi_{\text{strain}} = - \frac{v_{\rm eff}^2 e^2}{6\pi} \frac{\tanh\left(\frac{m}{2 k_B T}\right)}{m}.
\end{equation}
In the massless limit $m\rightarrow 0$, Eq.~\eqref{eq:suscsl} reduces to the well-known $1/T$ dependence: $\chi_{\text{strain}}|_{_{m=0}} = - \frac{v_{\rm eff}^2 e^2}{12\pi k_B T}$. Our numerical simulations quantitatively confirm the hyperbolic tangent dependence on inverse temperature of this ``diamagnetic" strain response in both KSL and CSL phases, as shown in Fig.~\ref{fig3}(a)-(b). In addition to the low-energy Dirac contributions, there are ``paramagnetic" responses from higher-energy modes, which become significant at higher temperatures. A detailed discussion of these high-energy non-Dirac contributions is provided in the SM. 

Notably, this low-temperature diamagnetic behavior in KSL and CSL phases is observable even in very small systems (e.g., a  lattice with $4\times 2$ unit cells), and can be directly captured through exact diagonalization (ED) of the spin model in Eq.\eqref{eq:H}. As shown in Fig.~\ref{fig3}(c), even with the contribution from $\mathbb{Z}_2$ fluxes, the Majorana-induced diamagnetic  strain susceptibility and its $1/T$ dependence persist. Meanwhile it is straightforward to extract that the response of $\mathbb{Z}_2$ fluxes to strain is paramagnetic [see SM]. If we deduct the contribution from $\mathbb{Z}_2$ fluxes by introducing a penalty term:
\begin{equation}
H_{\text{penalty}}=-\mu \hat{W}_{p}
\end{equation}
where $\hat{W}_{p}$ that is the product of six link operators $\hat{u}_{ij}$ that belong to a hexagon, and taking a large $\mu$, then the ED results from the spin model match perfectly with those from the Majorana hopping models~\footnote{Beyond serving as a consistency check, this calculation on small system sizes suggests that the proposed diamagnetic strain susceptibility could be experimentally probed in Kitaev spin liquids realized in cold-atom systems, where system sizes are typically small.}.

In contrast, the strain response in the Majorana metal phase is qualitatively distinct. As shown in Fig.~\ref{fig3}(d), ED calculations of the spin model with $h=0.6$ on a $4 \times 2$ unit-cell cluster—deep in the Majorana metal phase—reveal that the strain susceptibility is ``paramagnetic" and exhibits a different temperature dependence. This observed paramagnetic behavior is consistent with the absence of pseudo-Landau levels in the Majorana metal phase, as discussed in the previous section. Further details of the ED calculations and the corresponding Majorana-hopping model calculations can be found in the SM.

\begin{figure}[h!]
    \centering
    \includegraphics[width=0.98\linewidth]{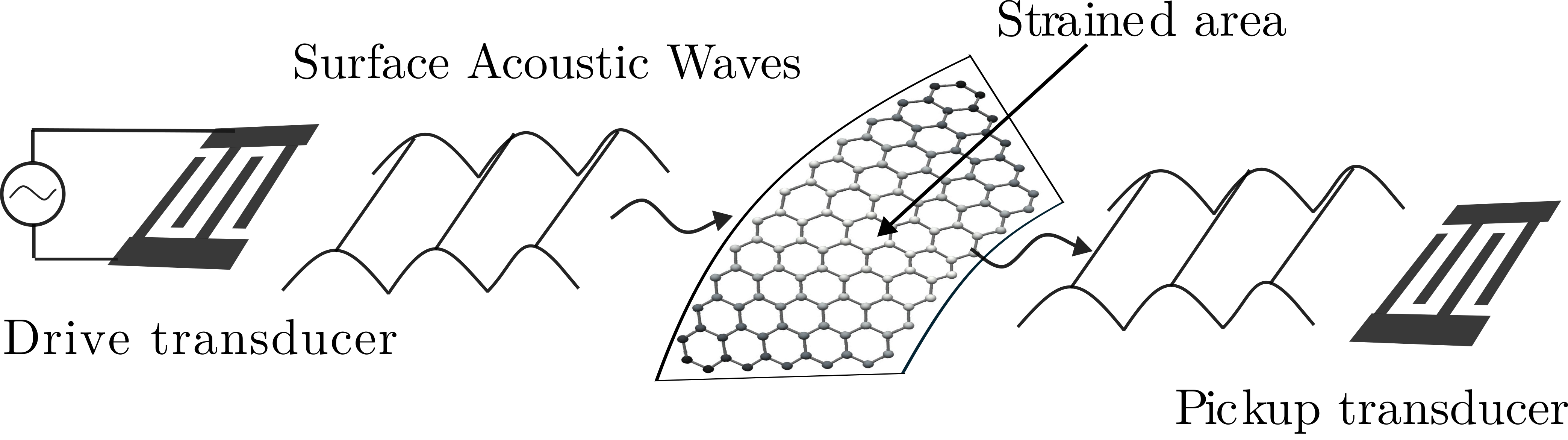}
    \caption{Schematic of a LRUS measurement using focused surface acoustic waves. A focused surface acoustic wave beam is launched from the left transducer, directed through a strained graphene region on a surface acoustic wave-supporting substrate, and detected by the right transducer. Inter–pseudo-Landau-level excitations induced by strain lead to resonant absorption, visible as pronounced features in the transmitted surface acoustic wave attenuation.}
    \label{fig:RUS-probe}
\end{figure}

\noindent\magenta{\it Probe pseudo-Landau levels and strain susceptibility using local resonant ultrasound spectroscopy.|} The Landau levels and strain susceptibility discussed above are experimentally accessible through local resonant ultrasound spectroscopy (RUS) measurements, which has previously been proposed as a probe to detect spinon Fermi surfaces\cite{ZhouLee2011, SerbynLee2013}.  RUS~\cite{dukhin2002,ramshaw2015,balakirev2019,hauspurg2024,Theuss2024_2} is a technique to measure the elastic moduli of a sample, but it works best in unstrained bulk three-dimensional samples. However, the quantum spin-liquid phases studied here typically emerge in effectively two-dimensional thin-film geometries (e.g. thin films of $\alpha$-RuCl$_3$) and exhibit strain that is highly localized on sub-micron length scales. Therefore, we need a local RUS (LRUS) technique \cite{Rus_local_2020} in 2D to spatially resolve the effects of strain. This can be done by probing the system with a focused surface acoustic wave used for thin films~\cite{hurley2001}, as illustrated in Fig.~\ref{fig:RUS-probe}.

Let us first demonstrate that the strain-induced pseudo-Landau levels in both the KSL and the CSL phases can be probed using LRUS. In a typical RUS setup, two piezoelectric transducers are attached to the sample: one generates ultrasonic waves, while the other detects the resulting lattice vibrations—essentially applying dynamical strain across the two sides [see Fig.~\ref{fig:RUS-probe}].
The absorption of lattice vibrations via coupling between Majoranas and strain leads to sound attenuation, characterized by the attenuation coefficient
\begin{equation}
\alpha\propto \frac{\Gamma}{v_s},
\end{equation}
where $\Gamma$ is the absorption rate, and $v_{s}$ is the sound velocity. In the following, we demonstrate that: (a) the energy spacing between pseudo-Landau levels lies within the frequency range accessible by LRUS experiments; and (b) our estimate shows that the absorption rate $\Gamma$ is significant and should give an observable sound attenuation coefficient $\alpha$.

To see point (a), we estimate the gap between pseudo-Landau levels in the KSL phases:
\begin{equation}
\label{eq:landaugap}
\Delta_{n}=2\sqrt{2}(\sqrt{n+1}-\sqrt{n})\hbar v_{\text{eff}}\sqrt{\frac{\beta \mathcal{S}}{aL}},
\end{equation}
where $\hbar v_{\text{eff}}=3J_0a \approx 18 \text{meV\AA}$ with $J_{0}=1.75 \text{meV}$ \footnote{Note that the $J_{0}$ defined in this work is one fourth of those defiend in ~\cite{li2021,hauspurg2024}} and  $a\approx 3.5 \text{\AA}$ in $\alpha$-RuCl$_3$. Given a strain strength such that $\beta\mathcal{S}=0.1$, and a typical linear size of the $\alpha$-RuCl$_3$ samples $L\sim 1 \text{mm}$, the gap between the zeroth and first pseudo-Landau level can be estimated as
\begin{equation}
\Delta_{0}\approx 51 \text{meV\AA}\times 5.3\times10^{-5}\text{\AA}^{-1}\approx 2.7\mu\text{eV}, 
\end{equation}
which is approximately $650 \text{MHz}$ and falls into the accessible range of $1\text{MHz}$ to $1\text{GHz}$~\footnote{Actually, Ref.~\cite{hauspurg2024} conducted the RUS measurement in $\alpha$-RuCl$_3$ and scan the frequency from $20\text{MHz}$ to $120 \text{MHz}$. It is quite promising to see the strain induced pseudo-Landau levels using LRUS proposed by us.}.

Next, let us demonstrate point (b). Since we know that at low energy the strain couples to the Dirac modes like an effective vector gauge potential, we focus on a low energy Hamiltonian 
\begin{equation}
H=v_{\rm eff}(\mathbf{p}+e\mathbf{A}_{\text{eff},0}(\mathbf{r}))\cdot \boldsymbol{\sigma}+v_{\rm eff}e\mathbf{A}_{\text{eff}}(\mathbf{r},t)\cdot\boldsymbol{\sigma},
\end{equation}
where $\mathbf{A}_{\text{eff},0}$ and $\mathbf{A}_{\text{eff}}(\mathbf{r},t)$ are the gauge fields generated by static and dynamical strains, respectively. We treat the last term as a time-dependent perturbation, $\mathbf{A_{\rm eff}}(\mathbf{r},t)=\int d\mathbf{q}/(2\pi)^2e^{i\mathbf{q}\cdot \mathbf{r}}[\mathcal{A}_{\mathbf{q}}e^{-i\omega_{\mathbf{q}}t}+\mathcal{A}^{\ast}_{\mathbf{q}}e^{i\omega_{\mathbf{q}}t}]$. The absorption rate can be generally evaluated through the Fermi Golden rule,
\begin{equation}
\Gamma_{\mathbf{q}}=\frac{2\pi}{\hbar}\sum_{f,i}|\bra{\phi_{f}}e^{i\mathbf{q}\cdot\mathbf{r}}v_{\rm eff}e\mathcal{A}_{\mathbf{q}}\cdot\sigma\ket{\phi_{i}}|^2\delta(\epsilon_{i}-\epsilon_{f}-\hbar\omega_{\mathbf{q}}),
\end{equation}
where $\ket{\phi_{f/i}}$ are eigenstates of $H_{0}=v_{\rm eff}(\mathbf{p}+e\mathbf{A}_{\text{eff},0}(\mathbf{r}))\cdot \boldsymbol{\sigma}$. If $f$ and $i$ are in the first and zeroth pseudo-Landau levels, then when $\hbar\omega_{\mathbf{q}}=\Delta_0$, there will be a divergent absorption rate due to the divergent DOS, which results in a significant sound attenuation. 
Note that in realistic systems, effects like lattice regularization and the broadening of pseudo-Landau levels will always regulate the divergent absorption rate into a finite value.

Let us next consider the simplest case where $\mathbf{A}(\mathbf{r},t)$ only has a non-zero $x$-component. Under the Landau gauge, the eigenstates of $H_{0}$ in the zeroth and first pseudo-Landau levels can be represented as
\begin{equation}
\ket{\phi_{0}}=\begin{pmatrix}
    \ket{0,k_{y}}
    \\
    0
\end{pmatrix}, \ \ket{\phi_{1}}=\frac{\sqrt{2}}{2}\begin{pmatrix}
    \ket{1,k_{y}}
    \\
    \ket{0,k_{y}}
\end{pmatrix},
\end{equation}
where $\ket{n,k_{y}}$ represent the states under Landau gauge in the $n$-th Landau level of 2D electron gas. Then the matrix element $v_{\rm eff}q\mathcal{A}_{\mathbf{q}}\bra{\phi_{1}}e^{i\mathbf{q}\cdot\mathbf{r}}\sigma_{x}\ket{\phi_{0}}\sim \int d\mathbf{r}e^{i\mathbf{q}\cdot\mathbf{r}}\psi^{\ast}_{0,k_{y}'}\psi_{0,k_{y}}$ which is not zero when $(0,k_{y}')=\mathbf{q}+(0,k_{y})$, where $\psi_{0,k_{y}}=\frac{1}{\sqrt{\pi^{1/2}L_{y}l_B}}e^{ik_{y}y}e^{-(x-k_{y}l_{B}^2)^2/2l_{B}^2}$ is the stripe-like wavefunction of the lowest Landau level. In conclusion, the absorption of dynamical strain mode is not forbidden by any selection rules and can have significant rate due to the large DOS of the pseudo-Landau levels.

As for the measurements of strain susceptibility, LRUS can generally measure elements of the elastic tensor \cite{Zadler2004, Tennakoon2017, GOODLET2021140507}, of which the contribution from electron spins is the strain susceptibility [see Eq.\eqref{eq:elasticity}].
The elements of the elastic tensor or the stiffness arise from a contribution of the lattice, as well as the contribution of electron spins, which we calculated in this paper. Since the lattice contribution to the elastic tensor is generally insensitive to external magnetic fields, the field-dependent contrast in the elastic response between the diamagnetic KSL/CSL phases and the paramagnetic Majorana metal phases naturally isolates the stiffness contributed by spins. In particular, a decrease in stiffness (i.e., magnitude of elastic tensor) is expected when the external magnetic field increases and drives the phase transition from the CSL to the Majorana metal phase, as the strain susceptibility arising from spins changes sign, from negative (diamagnetic), to positive (paramagnetic), whereas the lattice contribution to the strain susceptibility is always diamagnetic. The spin contribution to the strain susceptibility may have a maximum influence on the experimental results in the KSL phase at a very low temperature, as there this quantity is expected to diverge.
Furthermore, a direct comparison of the magnetic field dependence of the total strain susceptibility between a magnetic compound (e.g., $\alpha$-RuCl$_3$) and its non-magnetic structural analog (which can be obtained by replacing Ru with Rh in  $\alpha$-RuCl$_3$) would allow one to disentangle the phononic contribution and thereby quantify the strain susceptibility associated with spin degrees of freedom. Even if the strain is localized to a small region [see Fig.\ref{fig:RUS-probe}], its effects on strain susceptibility can be experimentally detected by probing the highly strained region with a focused acoustic beam.

\noindent\magenta{\it Conclusion and remarks.|} In summary, we demonstrate that strain responses provide a useful probe for fractionalization in quantum spin liquid phases. We show that the three quantum spin liquid phases of the Kitaev model under an out-of-plane magnetic field exhibit distinct strain susceptibilities, leading to corresponding changes in the elastic coefficients with magnetic field that can be measured experimentally. A characteristic negative strain susceptibility emerges at low magnetic fields, signaling the formation of pseudo-Landau levels, while a positive  strain susceptibility dominates at moderate fields in the Majorana metal phase. Given the energy scale associated with these pseudo-Landau levels at low magnetic fields, we propose that measuring sound attenuation in LRUS experiments offers a promising experimental approach to detect them and thereby reveal the underlying fractionalized excitations. Furthermore, the strain susceptibility investigated in our work corresponds to the spin contribution to the elastic tensor, whose magnetic-field dependence can be accessed through LRUS measurements and serves as a sensitive probe of the distinct properties of charge-neutral fractionalized excitations across different QSL phases. We hope that our results will stimulate future LRUS experiments aimed at detecting strain responses of fractionalized excitations in frustrated magnetic systems.


\begin{acknowledgments}
\textbf{Acknowledgments:}
We thank Brad Ramshaw for fruitful discussions. P.Z. was primarily supported by the Center for Emergent Materials, an NSF MRSEC, under award number DMR-2011876, NT acknowledges support from the National Science Foundation Division of Mathematical Sciences Grant No. 2138905.
\end{acknowledgments}

\bibliography{refs.bib}

\begin{thebibliography}{44}%
\makeatletter
\providecommand \@ifxundefined [1]{%
 \@ifx{#1\undefined}
}%
\providecommand \@ifnum [1]{%
 \ifnum #1\expandafter \@firstoftwo
 \else \expandafter \@secondoftwo
 \fi
}%
\providecommand \@ifx [1]{%
 \ifx #1\expandafter \@firstoftwo
 \else \expandafter \@secondoftwo
 \fi
}%
\providecommand \natexlab [1]{#1}%
\providecommand \enquote  [1]{``#1''}%
\providecommand \bibnamefont  [1]{#1}%
\providecommand \bibfnamefont [1]{#1}%
\providecommand \citenamefont [1]{#1}%
\providecommand \href@noop [0]{\@secondoftwo}%
\providecommand \href [0]{\begingroup \@sanitize@url \@href}%
\providecommand \@href[1]{\@@startlink{#1}\@@href}%
\providecommand \@@href[1]{\endgroup#1\@@endlink}%
\providecommand \@sanitize@url [0]{\catcode `\\12\catcode `\$12\catcode `\&12\catcode `\#12\catcode `\^12\catcode `\_12\catcode `\%12\relax}%
\providecommand \@@startlink[1]{}%
\providecommand \@@endlink[0]{}%
\providecommand \url  [0]{\begingroup\@sanitize@url \@url }%
\providecommand \@url [1]{\endgroup\@href {#1}{\urlprefix }}%
\providecommand \urlprefix  [0]{URL }%
\providecommand \Eprint [0]{\href }%
\providecommand \doibase [0]{https://doi.org/}%
\providecommand \selectlanguage [0]{\@gobble}%
\providecommand \bibinfo  [0]{\@secondoftwo}%
\providecommand \bibfield  [0]{\@secondoftwo}%
\providecommand \translation [1]{[#1]}%
\providecommand \BibitemOpen [0]{}%
\providecommand \bibitemStop [0]{}%
\providecommand \bibitemNoStop [0]{.\EOS\space}%
\providecommand \EOS [0]{\spacefactor3000\relax}%
\providecommand \BibitemShut  [1]{\csname bibitem#1\endcsname}%
\let\auto@bib@innerbib\@empty
\bibitem [{\citenamefont {Han}\ \emph {et~al.}(2012)\citenamefont {Han}, \citenamefont {Helton}, \citenamefont {Chu}, \citenamefont {Nocera}, \citenamefont {Rodriguez-Rivera}, \citenamefont {Broholm},\ and\ \citenamefont {Lee}}]{han2012}%
  \BibitemOpen
  \bibfield  {author} {\bibinfo {author} {\bibfnamefont {T.-H.}\ \bibnamefont {Han}}, \bibinfo {author} {\bibfnamefont {J.~S.}\ \bibnamefont {Helton}}, \bibinfo {author} {\bibfnamefont {S.}~\bibnamefont {Chu}}, \bibinfo {author} {\bibfnamefont {D.~G.}\ \bibnamefont {Nocera}}, \bibinfo {author} {\bibfnamefont {J.~A.}\ \bibnamefont {Rodriguez-Rivera}}, \bibinfo {author} {\bibfnamefont {C.}~\bibnamefont {Broholm}},\ and\ \bibinfo {author} {\bibfnamefont {Y.~S.}\ \bibnamefont {Lee}},\ }\bibfield  {title} {\bibinfo {title} {Fractionalized excitations in the spin-liquid state of a kagome-lattice antiferromagnet},\ }\href@noop {} {\bibfield  {journal} {\bibinfo  {journal} {Nature}\ }\textbf {\bibinfo {volume} {492}},\ \bibinfo {pages} {406} (\bibinfo {year} {2012})}\BibitemShut {NoStop}%
\bibitem [{\citenamefont {Banerjee}\ \emph {et~al.}(2016)\citenamefont {Banerjee}, \citenamefont {Bridges}, \citenamefont {Yan}, \citenamefont {Aczel}, \citenamefont {Li}, \citenamefont {Stone}, \citenamefont {Granroth}, \citenamefont {Lumsden}, \citenamefont {Yiu}, \citenamefont {Knolle} \emph {et~al.}}]{banerjee2016}%
  \BibitemOpen
  \bibfield  {author} {\bibinfo {author} {\bibfnamefont {A.}~\bibnamefont {Banerjee}}, \bibinfo {author} {\bibfnamefont {C.}~\bibnamefont {Bridges}}, \bibinfo {author} {\bibfnamefont {J.-Q.}\ \bibnamefont {Yan}}, \bibinfo {author} {\bibfnamefont {A.}~\bibnamefont {Aczel}}, \bibinfo {author} {\bibfnamefont {L.}~\bibnamefont {Li}}, \bibinfo {author} {\bibfnamefont {M.}~\bibnamefont {Stone}}, \bibinfo {author} {\bibfnamefont {G.}~\bibnamefont {Granroth}}, \bibinfo {author} {\bibfnamefont {M.}~\bibnamefont {Lumsden}}, \bibinfo {author} {\bibfnamefont {Y.}~\bibnamefont {Yiu}}, \bibinfo {author} {\bibfnamefont {J.}~\bibnamefont {Knolle}}, \emph {et~al.},\ }\bibfield  {title} {\bibinfo {title} {Proximate {Kitaev} quantum spin liquid behaviour in a honeycomb magnet},\ }\href {https://doi.org/10.1038/nmat4604} {\bibfield  {journal} {\bibinfo  {journal} {Nature materials}\ }\textbf {\bibinfo {volume} {15}},\ \bibinfo {pages} {733} (\bibinfo {year} {2016})}\BibitemShut {NoStop}%
\bibitem [{\citenamefont {Wulferding}\ \emph {et~al.}(2010)\citenamefont {Wulferding}, \citenamefont {Lemmens}, \citenamefont {Scheib}, \citenamefont {R\"oder}, \citenamefont {Mendels}, \citenamefont {Chu}, \citenamefont {Han},\ and\ \citenamefont {Lee}}]{wulferding2010}%
  \BibitemOpen
  \bibfield  {author} {\bibinfo {author} {\bibfnamefont {D.}~\bibnamefont {Wulferding}}, \bibinfo {author} {\bibfnamefont {P.}~\bibnamefont {Lemmens}}, \bibinfo {author} {\bibfnamefont {P.}~\bibnamefont {Scheib}}, \bibinfo {author} {\bibfnamefont {J.}~\bibnamefont {R\"oder}}, \bibinfo {author} {\bibfnamefont {P.}~\bibnamefont {Mendels}}, \bibinfo {author} {\bibfnamefont {S.}~\bibnamefont {Chu}}, \bibinfo {author} {\bibfnamefont {T.}~\bibnamefont {Han}},\ and\ \bibinfo {author} {\bibfnamefont {Y.~S.}\ \bibnamefont {Lee}},\ }\bibfield  {title} {\bibinfo {title} {Interplay of thermal and quantum spin fluctuations in the kagome lattice compound herbertsmithite},\ }\href {https://doi.org/10.1103/PhysRevB.82.144412} {\bibfield  {journal} {\bibinfo  {journal} {Phys. Rev. B}\ }\textbf {\bibinfo {volume} {82}},\ \bibinfo {pages} {144412} (\bibinfo {year} {2010})}\BibitemShut {NoStop}%
\bibitem [{\citenamefont {Gupta}\ \emph {et~al.}(2016)\citenamefont {Gupta}, \citenamefont {Sriluckshmy}, \citenamefont {Mehlawat}, \citenamefont {Balodhi}, \citenamefont {Mishra}, \citenamefont {Hassan}, \citenamefont {Ramakrishnan}, \citenamefont {Muthu}, \citenamefont {Singh},\ and\ \citenamefont {Sood}}]{gupta2016}%
  \BibitemOpen
  \bibfield  {author} {\bibinfo {author} {\bibfnamefont {S.~N.}\ \bibnamefont {Gupta}}, \bibinfo {author} {\bibfnamefont {P.}~\bibnamefont {Sriluckshmy}}, \bibinfo {author} {\bibfnamefont {K.}~\bibnamefont {Mehlawat}}, \bibinfo {author} {\bibfnamefont {A.}~\bibnamefont {Balodhi}}, \bibinfo {author} {\bibfnamefont {D.~K.}\ \bibnamefont {Mishra}}, \bibinfo {author} {\bibfnamefont {S.}~\bibnamefont {Hassan}}, \bibinfo {author} {\bibfnamefont {T.}~\bibnamefont {Ramakrishnan}}, \bibinfo {author} {\bibfnamefont {D.}~\bibnamefont {Muthu}}, \bibinfo {author} {\bibfnamefont {Y.}~\bibnamefont {Singh}},\ and\ \bibinfo {author} {\bibfnamefont {A.}~\bibnamefont {Sood}},\ }\bibfield  {title} {\bibinfo {title} {Raman signatures of strong {Kitaev} exchange correlations in {(Na$_{1- x}$Li$_x$)$_2$IrO$_3$}: {Experiments} and theory},\ }\href {https://iopscience.iop.org/article/10.1209/0295-5075/114/47004} {\bibfield  {journal} {\bibinfo  {journal} {Europhysics Letters}\ }\textbf {\bibinfo {volume} {114}},\ \bibinfo {pages}
  {47004} (\bibinfo {year} {2016})}\BibitemShut {NoStop}%
\bibitem [{\citenamefont {Kasahara}\ \emph {et~al.}(2018)\citenamefont {Kasahara}, \citenamefont {Ohnishi}, \citenamefont {Mizukami}, \citenamefont {Tanaka}, \citenamefont {Ma}, \citenamefont {Sugii}, \citenamefont {Kurita}, \citenamefont {Tanaka}, \citenamefont {Nasu}, \citenamefont {Motome} \emph {et~al.}}]{kasahara2018}%
  \BibitemOpen
  \bibfield  {author} {\bibinfo {author} {\bibfnamefont {Y.}~\bibnamefont {Kasahara}}, \bibinfo {author} {\bibfnamefont {T.}~\bibnamefont {Ohnishi}}, \bibinfo {author} {\bibfnamefont {Y.}~\bibnamefont {Mizukami}}, \bibinfo {author} {\bibfnamefont {O.}~\bibnamefont {Tanaka}}, \bibinfo {author} {\bibfnamefont {S.}~\bibnamefont {Ma}}, \bibinfo {author} {\bibfnamefont {K.}~\bibnamefont {Sugii}}, \bibinfo {author} {\bibfnamefont {N.}~\bibnamefont {Kurita}}, \bibinfo {author} {\bibfnamefont {H.}~\bibnamefont {Tanaka}}, \bibinfo {author} {\bibfnamefont {J.}~\bibnamefont {Nasu}}, \bibinfo {author} {\bibfnamefont {Y.}~\bibnamefont {Motome}}, \emph {et~al.},\ }\bibfield  {title} {\bibinfo {title} {{Majorana quantization and half-integer thermal quantum Hall effect in a Kitaev spin liquid}},\ }\href {https://doi.org/10.1038/s41586-018-0274-0} {\bibfield  {journal} {\bibinfo  {journal} {Nature}\ }\textbf {\bibinfo {volume} {559}},\ \bibinfo {pages} {227} (\bibinfo {year} {2018})}\BibitemShut {NoStop}%
\bibitem [{\citenamefont {Czajka}\ \emph {et~al.}(2021)\citenamefont {Czajka}, \citenamefont {Gao}, \citenamefont {Hirschberger}, \citenamefont {Lampen-Kelley}, \citenamefont {Banerjee}, \citenamefont {Yan}, \citenamefont {Mandrus}, \citenamefont {Nagler},\ and\ \citenamefont {Ong}}]{czajka2021}%
  \BibitemOpen
  \bibfield  {author} {\bibinfo {author} {\bibfnamefont {P.}~\bibnamefont {Czajka}}, \bibinfo {author} {\bibfnamefont {T.}~\bibnamefont {Gao}}, \bibinfo {author} {\bibfnamefont {M.}~\bibnamefont {Hirschberger}}, \bibinfo {author} {\bibfnamefont {P.}~\bibnamefont {Lampen-Kelley}}, \bibinfo {author} {\bibfnamefont {A.}~\bibnamefont {Banerjee}}, \bibinfo {author} {\bibfnamefont {J.}~\bibnamefont {Yan}}, \bibinfo {author} {\bibfnamefont {D.~G.}\ \bibnamefont {Mandrus}}, \bibinfo {author} {\bibfnamefont {S.~E.}\ \bibnamefont {Nagler}},\ and\ \bibinfo {author} {\bibfnamefont {N.}~\bibnamefont {Ong}},\ }\bibfield  {title} {\bibinfo {title} {{Oscillations of the thermal conductivity in the spin-liquid state of $\alpha$-Ru{Cl}$_3$}},\ }\href {https://doi.org/https://doi.org/10.1038/s41567-021-01243-x} {\bibfield  {journal} {\bibinfo  {journal} {Nature Physics}\ }\textbf {\bibinfo {volume} {17}},\ \bibinfo {pages} {915} (\bibinfo {year} {2021})}\BibitemShut {NoStop}%
\bibitem [{\citenamefont {Kohsaka}\ \emph {et~al.}(2024)\citenamefont {Kohsaka}, \citenamefont {Akutagawa}, \citenamefont {Omachi}, \citenamefont {Iwamichi}, \citenamefont {Ono}, \citenamefont {Tanaka}, \citenamefont {Tateishi}, \citenamefont {Murayama}, \citenamefont {Suetsugu}, \citenamefont {Hashimoto}, \citenamefont {Shibauchi}, \citenamefont {Takahashi}, \citenamefont {Nikolaev}, \citenamefont {Mizushima}, \citenamefont {Fujimoto}, \citenamefont {Terashima}, \citenamefont {Asaba}, \citenamefont {Kasahara},\ and\ \citenamefont {Matsuda}}]{Kohsaka2024}%
  \BibitemOpen
  \bibfield  {author} {\bibinfo {author} {\bibfnamefont {Y.}~\bibnamefont {Kohsaka}}, \bibinfo {author} {\bibfnamefont {S.}~\bibnamefont {Akutagawa}}, \bibinfo {author} {\bibfnamefont {S.}~\bibnamefont {Omachi}}, \bibinfo {author} {\bibfnamefont {Y.}~\bibnamefont {Iwamichi}}, \bibinfo {author} {\bibfnamefont {T.}~\bibnamefont {Ono}}, \bibinfo {author} {\bibfnamefont {I.}~\bibnamefont {Tanaka}}, \bibinfo {author} {\bibfnamefont {S.}~\bibnamefont {Tateishi}}, \bibinfo {author} {\bibfnamefont {H.}~\bibnamefont {Murayama}}, \bibinfo {author} {\bibfnamefont {S.}~\bibnamefont {Suetsugu}}, \bibinfo {author} {\bibfnamefont {K.}~\bibnamefont {Hashimoto}}, \bibinfo {author} {\bibfnamefont {T.}~\bibnamefont {Shibauchi}}, \bibinfo {author} {\bibfnamefont {M.~O.}\ \bibnamefont {Takahashi}}, \bibinfo {author} {\bibfnamefont {S.}~\bibnamefont {Nikolaev}}, \bibinfo {author} {\bibfnamefont {T.}~\bibnamefont {Mizushima}}, \bibinfo {author} {\bibfnamefont {S.}~\bibnamefont {Fujimoto}}, \bibinfo {author} {\bibfnamefont
  {T.}~\bibnamefont {Terashima}}, \bibinfo {author} {\bibfnamefont {T.}~\bibnamefont {Asaba}}, \bibinfo {author} {\bibfnamefont {Y.}~\bibnamefont {Kasahara}},\ and\ \bibinfo {author} {\bibfnamefont {Y.}~\bibnamefont {Matsuda}},\ }\bibfield  {title} {\bibinfo {title} {{Imaging Quantum Interference in a Monolayer Kitaev Quantum Spin Liquid Candidate}},\ }\href {https://doi.org/10.1103/PhysRevX.14.041026} {\bibfield  {journal} {\bibinfo  {journal} {Phys. Rev. X}\ }\textbf {\bibinfo {volume} {14}},\ \bibinfo {pages} {041026} (\bibinfo {year} {2024})}\BibitemShut {NoStop}%
\bibitem [{\citenamefont {Iordanskii}\ and\ \citenamefont {Koshelev}(1985)}]{IordanskiiKoshelev1985JETPLett}%
  \BibitemOpen
  \bibfield  {author} {\bibinfo {author} {\bibfnamefont {S.~V.}\ \bibnamefont {Iordanskii}}\ and\ \bibinfo {author} {\bibfnamefont {A.~E.}\ \bibnamefont {Koshelev}},\ }\bibfield  {title} {\bibinfo {title} {Dislocations and localization effects in multivalley conductors},\ }\href@noop {} {\bibfield  {journal} {\bibinfo  {journal} {JETP Letters}\ }\textbf {\bibinfo {volume} {41}},\ \bibinfo {pages} {574} (\bibinfo {year} {1985})},\ \bibinfo {note} {english translation of \emph{Pis'ma Zh. Eksp. Teor. Fiz.} \textbf{41}, 471 (1985).}\BibitemShut {Stop}%
\bibitem [{\citenamefont {Abanin}\ \emph {et~al.}(2007)\citenamefont {Abanin}, \citenamefont {Lee},\ and\ \citenamefont {Levitov}}]{Abanin2007}%
  \BibitemOpen
  \bibfield  {author} {\bibinfo {author} {\bibfnamefont {D.~A.}\ \bibnamefont {Abanin}}, \bibinfo {author} {\bibfnamefont {P.~A.}\ \bibnamefont {Lee}},\ and\ \bibinfo {author} {\bibfnamefont {L.~S.}\ \bibnamefont {Levitov}},\ }\bibfield  {title} {\bibinfo {title} {Randomness-induced $xy$ ordering in a graphene quantum hall ferromagnet},\ }\href {https://doi.org/10.1103/PhysRevLett.98.156801} {\bibfield  {journal} {\bibinfo  {journal} {Phys. Rev. Lett.}\ }\textbf {\bibinfo {volume} {98}},\ \bibinfo {pages} {156801} (\bibinfo {year} {2007})}\BibitemShut {NoStop}%
\bibitem [{\citenamefont {Guinea}\ \emph {et~al.}(2010)\citenamefont {Guinea}, \citenamefont {Katsnelson},\ and\ \citenamefont {Geim}}]{guinea2010}%
  \BibitemOpen
  \bibfield  {author} {\bibinfo {author} {\bibfnamefont {F.}~\bibnamefont {Guinea}}, \bibinfo {author} {\bibfnamefont {M.~I.}\ \bibnamefont {Katsnelson}},\ and\ \bibinfo {author} {\bibfnamefont {A.}~\bibnamefont {Geim}},\ }\bibfield  {title} {\bibinfo {title} {{Energy gaps and a zero-field quantum Hall effect in graphene by strain engineering}},\ }\href {https://doi.org/10.1038/nphys1420} {\bibfield  {journal} {\bibinfo  {journal} {Nature Physics}\ }\textbf {\bibinfo {volume} {6}},\ \bibinfo {pages} {30} (\bibinfo {year} {2010})}\BibitemShut {NoStop}%
\bibitem [{\citenamefont {Levy}\ \emph {et~al.}(2010)\citenamefont {Levy}, \citenamefont {Burke}, \citenamefont {Meaker}, \citenamefont {Panlasigui}, \citenamefont {Zettl}, \citenamefont {Guinea}, \citenamefont {Neto},\ and\ \citenamefont {Crommie}}]{levy2010}%
  \BibitemOpen
  \bibfield  {author} {\bibinfo {author} {\bibfnamefont {N.}~\bibnamefont {Levy}}, \bibinfo {author} {\bibfnamefont {S.~A.}\ \bibnamefont {Burke}}, \bibinfo {author} {\bibfnamefont {K.~L.}\ \bibnamefont {Meaker}}, \bibinfo {author} {\bibfnamefont {M.}~\bibnamefont {Panlasigui}}, \bibinfo {author} {\bibfnamefont {A.}~\bibnamefont {Zettl}}, \bibinfo {author} {\bibfnamefont {F.}~\bibnamefont {Guinea}}, \bibinfo {author} {\bibfnamefont {A.~H.~C.}\ \bibnamefont {Neto}},\ and\ \bibinfo {author} {\bibfnamefont {M.~F.}\ \bibnamefont {Crommie}},\ }\bibfield  {title} {\bibinfo {title} {Strain-induced pseudo–magnetic fields greater than 300 tesla in graphene nanobubbles},\ }\href {https://doi.org/10.1126/science.1191700} {\bibfield  {journal} {\bibinfo  {journal} {Science}\ }\textbf {\bibinfo {volume} {329}},\ \bibinfo {pages} {544} (\bibinfo {year} {2010})}\BibitemShut {NoStop}%
\bibitem [{\citenamefont {Berman}\ \emph {et~al.}(2022)\citenamefont {Berman}, \citenamefont {Kezerashvili}, \citenamefont {Lozovik},\ and\ \citenamefont {Ziegler}}]{Berman2022}%
  \BibitemOpen
  \bibfield  {author} {\bibinfo {author} {\bibfnamefont {O.~L.}\ \bibnamefont {Berman}}, \bibinfo {author} {\bibfnamefont {R.~Y.}\ \bibnamefont {Kezerashvili}}, \bibinfo {author} {\bibfnamefont {Y.~E.}\ \bibnamefont {Lozovik}},\ and\ \bibinfo {author} {\bibfnamefont {K.~G.}\ \bibnamefont {Ziegler}},\ }\bibfield  {title} {\bibinfo {title} {Strain-induced quantum hall phenomena of excitons in graphene},\ }\bibfield  {journal} {\bibinfo  {journal} {Scientific Reports}\ }\textbf {\bibinfo {volume} {12}},\ \href {https://doi.org/10.1038/s41598-022-06486-z} {10.1038/s41598-022-06486-z} (\bibinfo {year} {2022})\BibitemShut {NoStop}%
\bibitem [{\citenamefont {Schwarz}\ and\ \citenamefont {Vuorinen}(2000)}]{schwarz2000}%
  \BibitemOpen
  \bibfield  {author} {\bibinfo {author} {\bibfnamefont {R.}~\bibnamefont {Schwarz}}\ and\ \bibinfo {author} {\bibfnamefont {J.}~\bibnamefont {Vuorinen}},\ }\bibfield  {title} {\bibinfo {title} {Resonant ultrasound spectroscopy: applications, current status and limitations},\ }\href@noop {} {\bibfield  {journal} {\bibinfo  {journal} {Journal of Alloys and Compounds}\ }\textbf {\bibinfo {volume} {310}},\ \bibinfo {pages} {243} (\bibinfo {year} {2000})}\BibitemShut {NoStop}%
\bibitem [{\citenamefont {Dukhin}(2002)}]{dukhin2002}%
  \BibitemOpen
  \bibfield  {author} {\bibinfo {author} {\bibfnamefont {A.~S.}\ \bibnamefont {Dukhin}},\ }\href@noop {} {\emph {\bibinfo {title} {Ultrasound for Characterizing Colloids Particle Sizing, Zeta Potential Rheology}}}\ (\bibinfo  {publisher} {Elsevier},\ \bibinfo {year} {2002})\BibitemShut {NoStop}%
\bibitem [{\citenamefont {Balakirev}\ \emph {et~al.}(2019)\citenamefont {Balakirev}, \citenamefont {Ennaceur}, \citenamefont {Migliori}, \citenamefont {Maiorov},\ and\ \citenamefont {Migliori}}]{balakirev2019}%
  \BibitemOpen
  \bibfield  {author} {\bibinfo {author} {\bibfnamefont {F.~F.}\ \bibnamefont {Balakirev}}, \bibinfo {author} {\bibfnamefont {S.~M.}\ \bibnamefont {Ennaceur}}, \bibinfo {author} {\bibfnamefont {R.~J.}\ \bibnamefont {Migliori}}, \bibinfo {author} {\bibfnamefont {B.}~\bibnamefont {Maiorov}},\ and\ \bibinfo {author} {\bibfnamefont {A.}~\bibnamefont {Migliori}},\ }\bibfield  {title} {\bibinfo {title} {Resonant ultrasound spectroscopy: The essential toolbox},\ }\href {https://pubs.aip.org/aip/rsi/article/90/12/121401/1018068} {\bibfield  {journal} {\bibinfo  {journal} {Review of Scientific Instruments}\ }\textbf {\bibinfo {volume} {90}} (\bibinfo {year} {2019})}\BibitemShut {NoStop}%
\bibitem [{\citenamefont {Hauspurg}\ \emph {et~al.}(2024)\citenamefont {Hauspurg}, \citenamefont {Zherlitsyn}, \citenamefont {Helm}, \citenamefont {Felea}, \citenamefont {Wosnitza}, \citenamefont {Tsurkan}, \citenamefont {Choi}, \citenamefont {Do}, \citenamefont {Ye}, \citenamefont {Brenig},\ and\ \citenamefont {Perkins}}]{hauspurg2024}%
  \BibitemOpen
  \bibfield  {author} {\bibinfo {author} {\bibfnamefont {A.}~\bibnamefont {Hauspurg}}, \bibinfo {author} {\bibfnamefont {S.}~\bibnamefont {Zherlitsyn}}, \bibinfo {author} {\bibfnamefont {T.}~\bibnamefont {Helm}}, \bibinfo {author} {\bibfnamefont {V.}~\bibnamefont {Felea}}, \bibinfo {author} {\bibfnamefont {J.}~\bibnamefont {Wosnitza}}, \bibinfo {author} {\bibfnamefont {V.}~\bibnamefont {Tsurkan}}, \bibinfo {author} {\bibfnamefont {K.-Y.}\ \bibnamefont {Choi}}, \bibinfo {author} {\bibfnamefont {S.-H.}\ \bibnamefont {Do}}, \bibinfo {author} {\bibfnamefont {M.}~\bibnamefont {Ye}}, \bibinfo {author} {\bibfnamefont {W.}~\bibnamefont {Brenig}},\ and\ \bibinfo {author} {\bibfnamefont {N.~B.}\ \bibnamefont {Perkins}},\ }\bibfield  {title} {\bibinfo {title} {Fractionalized excitations probed by ultrasound},\ }\href {https://doi.org/10.1103/PhysRevB.109.144415} {\bibfield  {journal} {\bibinfo  {journal} {Phys. Rev. B}\ }\textbf {\bibinfo {volume} {109}},\ \bibinfo {pages} {144415} (\bibinfo {year} {2024})}\BibitemShut
  {NoStop}%
\bibitem [{\citenamefont {Theuss}\ \emph {et~al.}(2024)\citenamefont {Theuss}, \citenamefont {Simarro}, \citenamefont {Shragai}, \citenamefont {Grissonnanche}, \citenamefont {Hayes}, \citenamefont {Saha}, \citenamefont {Shishidou}, \citenamefont {Chen}, \citenamefont {Nakatsuji}, \citenamefont {Ran}, \citenamefont {Weinert}, \citenamefont {Butch}, \citenamefont {Paglione},\ and\ \citenamefont {Ramshaw}}]{Theuss2024_2}%
  \BibitemOpen
  \bibfield  {author} {\bibinfo {author} {\bibfnamefont {F.}~\bibnamefont {Theuss}}, \bibinfo {author} {\bibfnamefont {G.~d. l.~F.}\ \bibnamefont {Simarro}}, \bibinfo {author} {\bibfnamefont {A.}~\bibnamefont {Shragai}}, \bibinfo {author} {\bibfnamefont {G.}~\bibnamefont {Grissonnanche}}, \bibinfo {author} {\bibfnamefont {I.~M.}\ \bibnamefont {Hayes}}, \bibinfo {author} {\bibfnamefont {S.}~\bibnamefont {Saha}}, \bibinfo {author} {\bibfnamefont {T.}~\bibnamefont {Shishidou}}, \bibinfo {author} {\bibfnamefont {T.}~\bibnamefont {Chen}}, \bibinfo {author} {\bibfnamefont {S.}~\bibnamefont {Nakatsuji}}, \bibinfo {author} {\bibfnamefont {S.}~\bibnamefont {Ran}}, \bibinfo {author} {\bibfnamefont {M.}~\bibnamefont {Weinert}}, \bibinfo {author} {\bibfnamefont {N.~P.}\ \bibnamefont {Butch}}, \bibinfo {author} {\bibfnamefont {J.}~\bibnamefont {Paglione}},\ and\ \bibinfo {author} {\bibfnamefont {B.~J.}\ \bibnamefont {Ramshaw}},\ }\bibfield  {title} {\bibinfo {title} {Resonant ultrasound spectroscopy for irregularly shaped
  samples and its application to uranium ditelluride},\ }\href {https://doi.org/10.1103/PhysRevLett.132.066003} {\bibfield  {journal} {\bibinfo  {journal} {Phys. Rev. Lett.}\ }\textbf {\bibinfo {volume} {132}},\ \bibinfo {pages} {066003} (\bibinfo {year} {2024})}\BibitemShut {NoStop}%
\bibitem [{\citenamefont {{K}itaev}(2006)}]{kitaev2006anyons}%
  \BibitemOpen
  \bibfield  {author} {\bibinfo {author} {\bibfnamefont {A.}~\bibnamefont {{K}itaev}},\ }\bibfield  {title} {\bibinfo {title} {Anyons in an exactly solved model and beyond},\ }\href {https://doi.org/https://doi.org/10.1016/j.aop.2005.10.005} {\bibfield  {journal} {\bibinfo  {journal} {Annals of Physics}\ }\textbf {\bibinfo {volume} {321}},\ \bibinfo {pages} {2} (\bibinfo {year} {2006})}\BibitemShut {NoStop}%
\bibitem [{\citenamefont {Wang}\ \emph {et~al.}(2025)\citenamefont {Wang}, \citenamefont {Feng}, \citenamefont {Zhu}, \citenamefont {Chi}, \citenamefont {Liao}, \citenamefont {Trivedi},\ and\ \citenamefont {Xiang}}]{wang2025}%
  \BibitemOpen
  \bibfield  {author} {\bibinfo {author} {\bibfnamefont {K.}~\bibnamefont {Wang}}, \bibinfo {author} {\bibfnamefont {S.}~\bibnamefont {Feng}}, \bibinfo {author} {\bibfnamefont {P.}~\bibnamefont {Zhu}}, \bibinfo {author} {\bibfnamefont {R.}~\bibnamefont {Chi}}, \bibinfo {author} {\bibfnamefont {H.-J.}\ \bibnamefont {Liao}}, \bibinfo {author} {\bibfnamefont {N.}~\bibnamefont {Trivedi}},\ and\ \bibinfo {author} {\bibfnamefont {T.}~\bibnamefont {Xiang}},\ }\bibfield  {title} {\bibinfo {title} {Fractionalization signatures in the dynamics of quantum spin liquids},\ }\href {https://doi.org/10.1103/PhysRevB.111.L100402} {\bibfield  {journal} {\bibinfo  {journal} {Phys. Rev. B}\ }\textbf {\bibinfo {volume} {111}},\ \bibinfo {pages} {L100402} (\bibinfo {year} {2025})}\BibitemShut {NoStop}%
\bibitem [{\citenamefont {Zhu}\ \emph {et~al.}(2025)\citenamefont {Zhu}, \citenamefont {Feng}, \citenamefont {Wang}, \citenamefont {Xiang},\ and\ \citenamefont {Trivedi}}]{zhu2025}%
  \BibitemOpen
  \bibfield  {author} {\bibinfo {author} {\bibfnamefont {P.}~\bibnamefont {Zhu}}, \bibinfo {author} {\bibfnamefont {S.}~\bibnamefont {Feng}}, \bibinfo {author} {\bibfnamefont {K.}~\bibnamefont {Wang}}, \bibinfo {author} {\bibfnamefont {T.}~\bibnamefont {Xiang}},\ and\ \bibinfo {author} {\bibfnamefont {N.}~\bibnamefont {Trivedi}},\ }\bibfield  {title} {\bibinfo {title} {{Emergent quantum Majorana metal from a chiral spin liquid}},\ }\href {https://doi.org/10.1038/s41467-025-56789-8} {\bibfield  {journal} {\bibinfo  {journal} {Nature Communications}\ }\textbf {\bibinfo {volume} {16}},\ \bibinfo {pages} {2420} (\bibinfo {year} {2025})}\BibitemShut {NoStop}%
\bibitem [{\citenamefont {Feng}\ \emph {et~al.}(2025)\citenamefont {Feng}, \citenamefont {Zhu}, \citenamefont {Knolle},\ and\ \citenamefont {Knap}}]{feng2025}%
  \BibitemOpen
  \bibfield  {author} {\bibinfo {author} {\bibfnamefont {S.}~\bibnamefont {Feng}}, \bibinfo {author} {\bibfnamefont {P.}~\bibnamefont {Zhu}}, \bibinfo {author} {\bibfnamefont {J.}~\bibnamefont {Knolle}},\ and\ \bibinfo {author} {\bibfnamefont {M.}~\bibnamefont {Knap}},\ }\bibfield  {title} {\bibinfo {title} {Transient localization from fractionalization: vanishingly small heat conductivity in gapless quantum magnets},\ }\href@noop {} {\bibfield  {journal} {\bibinfo  {journal} {arXiv preprint arXiv:2509.07062}\ } (\bibinfo {year} {2025})}\BibitemShut {NoStop}%
\bibitem [{\citenamefont {Ribeiro}\ \emph {et~al.}(2009)\citenamefont {Ribeiro}, \citenamefont {Pereira}, \citenamefont {Peres}, \citenamefont {Briddon},\ and\ \citenamefont {Castro~Neto}}]{Ribeiro_strained_2009}%
  \BibitemOpen
  \bibfield  {author} {\bibinfo {author} {\bibfnamefont {R.~M.}\ \bibnamefont {Ribeiro}}, \bibinfo {author} {\bibfnamefont {V.~M.}\ \bibnamefont {Pereira}}, \bibinfo {author} {\bibfnamefont {N.~M.~R.}\ \bibnamefont {Peres}}, \bibinfo {author} {\bibfnamefont {P.~R.}\ \bibnamefont {Briddon}},\ and\ \bibinfo {author} {\bibfnamefont {A.~H.}\ \bibnamefont {Castro~Neto}},\ }\bibfield  {title} {\bibinfo {title} {Strained graphene: tight-binding and density functional calculations},\ }\href {https://doi.org/10.1088/1367-2630/11/11/115002} {\bibfield  {journal} {\bibinfo  {journal} {New Journal of Physics}\ }\textbf {\bibinfo {volume} {11}},\ \bibinfo {pages} {115002} (\bibinfo {year} {2009})}\BibitemShut {NoStop}%
\bibitem [{\citenamefont {Pearce}\ \emph {et~al.}(2016)\citenamefont {Pearce}, \citenamefont {Mariani},\ and\ \citenamefont {Burkard}}]{Pearce_tight-binding_2016}%
  \BibitemOpen
  \bibfield  {author} {\bibinfo {author} {\bibfnamefont {A.~J.}\ \bibnamefont {Pearce}}, \bibinfo {author} {\bibfnamefont {E.}~\bibnamefont {Mariani}},\ and\ \bibinfo {author} {\bibfnamefont {G.}~\bibnamefont {Burkard}},\ }\bibfield  {title} {\bibinfo {title} {Tight-binding approach to strain and curvature in monolayer transition-metal dichalcogenides},\ }\bibfield  {journal} {\bibinfo  {journal} {Physical Review B}\ }\textbf {\bibinfo {volume} {94}},\ \href {https://doi.org/10.1103/physrevb.94.155416} {10.1103/physrevb.94.155416} (\bibinfo {year} {2016})\BibitemShut {NoStop}%
\bibitem [{\citenamefont {Suzuura}\ and\ \citenamefont {Ando}(2002)}]{suzuura2002}%
  \BibitemOpen
  \bibfield  {author} {\bibinfo {author} {\bibfnamefont {H.}~\bibnamefont {Suzuura}}\ and\ \bibinfo {author} {\bibfnamefont {T.}~\bibnamefont {Ando}},\ }\bibfield  {title} {\bibinfo {title} {Phonons and electron-phonon scattering in carbon nanotubes},\ }\href {https://doi.org/10.1103/PhysRevB.65.235412} {\bibfield  {journal} {\bibinfo  {journal} {Phys. Rev. B}\ }\textbf {\bibinfo {volume} {65}},\ \bibinfo {pages} {235412} (\bibinfo {year} {2002})}\BibitemShut {NoStop}%
\bibitem [{\citenamefont {Ma\~nes}(2007)}]{manes2007}%
  \BibitemOpen
  \bibfield  {author} {\bibinfo {author} {\bibfnamefont {J.~L.}\ \bibnamefont {Ma\~nes}},\ }\bibfield  {title} {\bibinfo {title} {Symmetry-based approach to electron-phonon interactions in graphene},\ }\href {https://doi.org/10.1103/PhysRevB.76.045430} {\bibfield  {journal} {\bibinfo  {journal} {Phys. Rev. B}\ }\textbf {\bibinfo {volume} {76}},\ \bibinfo {pages} {045430} (\bibinfo {year} {2007})}\BibitemShut {NoStop}%
\bibitem [{\citenamefont {Masir}\ \emph {et~al.}(2013)\citenamefont {Masir}, \citenamefont {Moldovan},\ and\ \citenamefont {Peeters}}]{masir2013}%
  \BibitemOpen
  \bibfield  {author} {\bibinfo {author} {\bibfnamefont {M.~R.}\ \bibnamefont {Masir}}, \bibinfo {author} {\bibfnamefont {D.}~\bibnamefont {Moldovan}},\ and\ \bibinfo {author} {\bibfnamefont {F.}~\bibnamefont {Peeters}},\ }\bibfield  {title} {\bibinfo {title} {{Pseudo magnetic field in strained graphene: Revisited}},\ }\href {https://doi.org/10.1016/j.ssc.2013.04.001} {\bibfield  {journal} {\bibinfo  {journal} {Solid state communications}\ }\textbf {\bibinfo {volume} {175}},\ \bibinfo {pages} {76} (\bibinfo {year} {2013})}\BibitemShut {NoStop}%
\bibitem [{\citenamefont {Yamada}\ and\ \citenamefont {Suga}(2023)}]{yamada2023}%
  \BibitemOpen
  \bibfield  {author} {\bibinfo {author} {\bibfnamefont {T.}~\bibnamefont {Yamada}}\ and\ \bibinfo {author} {\bibfnamefont {S.-i.}\ \bibnamefont {Suga}},\ }\bibfield  {title} {\bibinfo {title} {{Strain-induced Landau levels of Majorana fermions in an anisotropically interacting Kitaev model on a honeycomb lattice}},\ }\href {https://journals.jps.jp/doi/10.7566/JPSJ.92.114705} {\bibfield  {journal} {\bibinfo  {journal} {Journal of the Physical Society of Japan}\ }\textbf {\bibinfo {volume} {92}},\ \bibinfo {pages} {114705} (\bibinfo {year} {2023})}\BibitemShut {NoStop}%
\bibitem [{Note1()}]{Note1}%
  \BibitemOpen
  \bibinfo {note} {Details about this calculation can be found in SM.}\BibitemShut {Stop}%
\bibitem [{Note2()}]{Note2}%
  \BibitemOpen
  \bibinfo {note} {In general, when the Hamiltonian is not linear in momentum, the strain will generally not couple to the orbital motion like an effective vector gauge field, and can serve as a trivial perturbation as illustrated in SM.}\BibitemShut {Stop}%
\bibitem [{\citenamefont {G\'omez-Santos}\ and\ \citenamefont {Stauber}(2011)}]{gomez2011}%
  \BibitemOpen
  \bibfield  {author} {\bibinfo {author} {\bibfnamefont {G.}~\bibnamefont {G\'omez-Santos}}\ and\ \bibinfo {author} {\bibfnamefont {T.}~\bibnamefont {Stauber}},\ }\bibfield  {title} {\bibinfo {title} {{Measurable Lattice Effects on the Charge and Magnetic Response in Graphene}},\ }\href {https://doi.org/10.1103/PhysRevLett.106.045504} {\bibfield  {journal} {\bibinfo  {journal} {Phys. Rev. Lett.}\ }\textbf {\bibinfo {volume} {106}},\ \bibinfo {pages} {045504} (\bibinfo {year} {2011})}\BibitemShut {NoStop}%
\bibitem [{\citenamefont {Ominato}\ and\ \citenamefont {Koshino}(2013)}]{ominato2013}%
  \BibitemOpen
  \bibfield  {author} {\bibinfo {author} {\bibfnamefont {Y.}~\bibnamefont {Ominato}}\ and\ \bibinfo {author} {\bibfnamefont {M.}~\bibnamefont {Koshino}},\ }\bibfield  {title} {\bibinfo {title} {Orbital magnetism of graphene flakes},\ }\href {https://doi.org/10.1103/PhysRevB.87.115433} {\bibfield  {journal} {\bibinfo  {journal} {Phys. Rev. B}\ }\textbf {\bibinfo {volume} {87}},\ \bibinfo {pages} {115433} (\bibinfo {year} {2013})}\BibitemShut {NoStop}%
\bibitem [{\citenamefont {Vallejo~Bustamante}\ \emph {et~al.}(2021)\citenamefont {Vallejo~Bustamante}, \citenamefont {Wu}, \citenamefont {Fermon}, \citenamefont {Pannetier-Lecoeur}, \citenamefont {Wakamura}, \citenamefont {Watanabe}, \citenamefont {Taniguchi}, \citenamefont {Pellegrin}, \citenamefont {Bernard}, \citenamefont {Daddinounou} \emph {et~al.}}]{vallejo2021}%
  \BibitemOpen
  \bibfield  {author} {\bibinfo {author} {\bibfnamefont {J.}~\bibnamefont {Vallejo~Bustamante}}, \bibinfo {author} {\bibfnamefont {N.}~\bibnamefont {Wu}}, \bibinfo {author} {\bibfnamefont {C.}~\bibnamefont {Fermon}}, \bibinfo {author} {\bibfnamefont {M.}~\bibnamefont {Pannetier-Lecoeur}}, \bibinfo {author} {\bibfnamefont {T.}~\bibnamefont {Wakamura}}, \bibinfo {author} {\bibfnamefont {K.}~\bibnamefont {Watanabe}}, \bibinfo {author} {\bibfnamefont {T.}~\bibnamefont {Taniguchi}}, \bibinfo {author} {\bibfnamefont {T.}~\bibnamefont {Pellegrin}}, \bibinfo {author} {\bibfnamefont {A.}~\bibnamefont {Bernard}}, \bibinfo {author} {\bibfnamefont {S.}~\bibnamefont {Daddinounou}}, \emph {et~al.},\ }\bibfield  {title} {\bibinfo {title} {{Detection of graphene’s divergent orbital diamagnetism at the Dirac point}},\ }\href {https://www.science.org/doi/full/10.1126/science.abf9396} {\bibfield  {journal} {\bibinfo  {journal} {Science}\ }\textbf {\bibinfo {volume} {374}},\ \bibinfo {pages} {1399} (\bibinfo {year}
  {2021})}\BibitemShut {NoStop}%
\bibitem [{Note3()}]{Note3}%
  \BibitemOpen
  \bibinfo {note} {Beyond serving as a consistency check, this calculation on small system sizes suggests that the proposed diamagnetic strain susceptibility could be experimentally probed in Kitaev spin liquids realized in cold-atom systems, where system sizes are typically small.}\BibitemShut {Stop}%
\bibitem [{\citenamefont {Zhou}\ and\ \citenamefont {Lee}(2011)}]{ZhouLee2011}%
  \BibitemOpen
  \bibfield  {author} {\bibinfo {author} {\bibfnamefont {Y.}~\bibnamefont {Zhou}}\ and\ \bibinfo {author} {\bibfnamefont {P.~A.}\ \bibnamefont {Lee}},\ }\bibfield  {title} {\bibinfo {title} {Spinon phonon interaction and ultrasonic attenuation in quantum spin liquids},\ }\href {https://doi.org/10.1103/PhysRevLett.106.056402} {\bibfield  {journal} {\bibinfo  {journal} {Phys. Rev. Lett.}\ }\textbf {\bibinfo {volume} {106}},\ \bibinfo {pages} {056402} (\bibinfo {year} {2011})}\BibitemShut {NoStop}%
\bibitem [{\citenamefont {Serbyn}\ and\ \citenamefont {Lee}(2013)}]{SerbynLee2013}%
  \BibitemOpen
  \bibfield  {author} {\bibinfo {author} {\bibfnamefont {M.}~\bibnamefont {Serbyn}}\ and\ \bibinfo {author} {\bibfnamefont {P.~A.}\ \bibnamefont {Lee}},\ }\bibfield  {title} {\bibinfo {title} {Spinon-phonon interaction in algebraic spin liquids},\ }\href {https://doi.org/10.1103/PhysRevB.87.174424} {\bibfield  {journal} {\bibinfo  {journal} {Phys. Rev. B}\ }\textbf {\bibinfo {volume} {87}},\ \bibinfo {pages} {174424} (\bibinfo {year} {2013})}\BibitemShut {NoStop}%
\bibitem [{\citenamefont {Ramshaw}\ \emph {et~al.}(2015)\citenamefont {Ramshaw}, \citenamefont {Shekhter}, \citenamefont {McDonald}, \citenamefont {Betts}, \citenamefont {Mitchell}, \citenamefont {Tobash}, \citenamefont {Mielke}, \citenamefont {Bauer},\ and\ \citenamefont {Migliori}}]{ramshaw2015}%
  \BibitemOpen
  \bibfield  {author} {\bibinfo {author} {\bibfnamefont {B.}~\bibnamefont {Ramshaw}}, \bibinfo {author} {\bibfnamefont {A.}~\bibnamefont {Shekhter}}, \bibinfo {author} {\bibfnamefont {R.~D.}\ \bibnamefont {McDonald}}, \bibinfo {author} {\bibfnamefont {J.~B.}\ \bibnamefont {Betts}}, \bibinfo {author} {\bibfnamefont {J.}~\bibnamefont {Mitchell}}, \bibinfo {author} {\bibfnamefont {P.}~\bibnamefont {Tobash}}, \bibinfo {author} {\bibfnamefont {C.}~\bibnamefont {Mielke}}, \bibinfo {author} {\bibfnamefont {E.}~\bibnamefont {Bauer}},\ and\ \bibinfo {author} {\bibfnamefont {A.}~\bibnamefont {Migliori}},\ }\bibfield  {title} {\bibinfo {title} {Avoided valence transition in a plutonium superconductor},\ }\href {https://www.pnas.org/doi/abs/10.1073/pnas.1421174112} {\bibfield  {journal} {\bibinfo  {journal} {Proceedings of the National Academy of Sciences}\ }\textbf {\bibinfo {volume} {112}},\ \bibinfo {pages} {3285} (\bibinfo {year} {2015})}\BibitemShut {NoStop}%
\bibitem [{\citenamefont {Rus}\ and\ \citenamefont {Grosse}(2020)}]{Rus_local_2020}%
  \BibitemOpen
  \bibfield  {author} {\bibinfo {author} {\bibfnamefont {J.}~\bibnamefont {Rus}}\ and\ \bibinfo {author} {\bibfnamefont {C.~U.}\ \bibnamefont {Grosse}},\ }\bibfield  {title} {\bibinfo {title} {Local ultrasonic resonance spectroscopy: A demonstration on plate inspection},\ }\bibfield  {journal} {\bibinfo  {journal} {Journal of Nondestructive Evaluation}\ }\textbf {\bibinfo {volume} {39}},\ \href {https://doi.org/10.1007/s10921-020-00674-5} {10.1007/s10921-020-00674-5} (\bibinfo {year} {2020})\BibitemShut {NoStop}%
\bibitem [{\citenamefont {Hurley}\ \emph {et~al.}(2001)\citenamefont {Hurley}, \citenamefont {Tewary},\ and\ \citenamefont {Richards}}]{hurley2001}%
  \BibitemOpen
  \bibfield  {author} {\bibinfo {author} {\bibfnamefont {D.~C.}\ \bibnamefont {Hurley}}, \bibinfo {author} {\bibfnamefont {V.~K.}\ \bibnamefont {Tewary}},\ and\ \bibinfo {author} {\bibfnamefont {A.}~\bibnamefont {Richards}},\ }\bibfield  {title} {\bibinfo {title} {Surface acoustic wave methods to determine the anisotropic elasticproperties of thin films},\ }\href@noop {} {\bibfield  {journal} {\bibinfo  {journal} {Measurement Science and Technology}\ }\textbf {\bibinfo {volume} {12}},\ \bibinfo {pages} {1486} (\bibinfo {year} {2001})}\BibitemShut {NoStop}%
\bibitem [{Note4()}]{Note4}%
  \BibitemOpen
  \bibinfo {note} {Note that the $J_{0}$ defined in this work is one fourth of those defiend in ~\cite {li2021,hauspurg2024}}\BibitemShut {NoStop}%
\bibitem [{Note5()}]{Note5}%
  \BibitemOpen
  \bibinfo {note} {Actually, Ref.~\cite {hauspurg2024} conducted the RUS measurement in $\alpha $-RuCl$_3$ and scan the frequency from $20\protect \text {MHz}$ to $120 \protect \text {MHz}$. It is quite promising to see the strain induced pseudo-Landau levels using LRUS proposed by us.}\BibitemShut {Stop}%
\bibitem [{\citenamefont {Zadler}\ \emph {et~al.}(2004)\citenamefont {Zadler}, \citenamefont {Le~Rousseau}, \citenamefont {Scales},\ and\ \citenamefont {Smith}}]{Zadler2004}%
  \BibitemOpen
  \bibfield  {author} {\bibinfo {author} {\bibfnamefont {B.~J.}\ \bibnamefont {Zadler}}, \bibinfo {author} {\bibfnamefont {J.~H.~L.}\ \bibnamefont {Le~Rousseau}}, \bibinfo {author} {\bibfnamefont {J.~A.}\ \bibnamefont {Scales}},\ and\ \bibinfo {author} {\bibfnamefont {M.~L.}\ \bibnamefont {Smith}},\ }\bibfield  {title} {\bibinfo {title} {Resonant ultrasound spectroscopy: theory and application},\ }\href {https://doi.org/10.1111/j.1365-246X.2004.02093.x} {\bibfield  {journal} {\bibinfo  {journal} {Geophysical Journal International}\ }\textbf {\bibinfo {volume} {156}},\ \bibinfo {pages} {154} (\bibinfo {year} {2004})},\ \Eprint {https://arxiv.org/abs/https://academic.oup.com/gji/article-pdf/156/1/154/6039715/156-1-154.pdf} {https://academic.oup.com/gji/article-pdf/156/1/154/6039715/156-1-154.pdf} \BibitemShut {NoStop}%
\bibitem [{\citenamefont {Tennakoon}\ \emph {et~al.}(2017)\citenamefont {Tennakoon}, \citenamefont {Gladden}, \citenamefont {Mookherjee}, \citenamefont {Besara},\ and\ \citenamefont {Siegrist}}]{Tennakoon2017}%
  \BibitemOpen
  \bibfield  {author} {\bibinfo {author} {\bibfnamefont {S.}~\bibnamefont {Tennakoon}}, \bibinfo {author} {\bibfnamefont {J.}~\bibnamefont {Gladden}}, \bibinfo {author} {\bibfnamefont {M.}~\bibnamefont {Mookherjee}}, \bibinfo {author} {\bibfnamefont {T.}~\bibnamefont {Besara}},\ and\ \bibinfo {author} {\bibfnamefont {T.}~\bibnamefont {Siegrist}},\ }\bibfield  {title} {\bibinfo {title} {Temperature-dependent elasticity of $\mathrm{Pb}[{(\mathrm{M}{\mathrm{g}}_{0.33}\mathrm{N}{\mathrm{b}}_{0.67})}_{1\text{\ensuremath{-}}x}\mathrm{T}{\mathrm{i}}_{x}]{\mathrm{o}}_{3}$},\ }\href {https://doi.org/10.1103/PhysRevB.96.134108} {\bibfield  {journal} {\bibinfo  {journal} {Phys. Rev. B}\ }\textbf {\bibinfo {volume} {96}},\ \bibinfo {pages} {134108} (\bibinfo {year} {2017})}\BibitemShut {NoStop}%
\bibitem [{\citenamefont {Goodlet}\ \emph {et~al.}(2021)\citenamefont {Goodlet}, \citenamefont {Murray}, \citenamefont {Bales}, \citenamefont {Rossin}, \citenamefont {Torbet},\ and\ \citenamefont {Pollock}}]{GOODLET2021140507}%
  \BibitemOpen
  \bibfield  {author} {\bibinfo {author} {\bibfnamefont {B.~R.}\ \bibnamefont {Goodlet}}, \bibinfo {author} {\bibfnamefont {S.~P.}\ \bibnamefont {Murray}}, \bibinfo {author} {\bibfnamefont {B.}~\bibnamefont {Bales}}, \bibinfo {author} {\bibfnamefont {J.}~\bibnamefont {Rossin}}, \bibinfo {author} {\bibfnamefont {C.~J.}\ \bibnamefont {Torbet}},\ and\ \bibinfo {author} {\bibfnamefont {T.~M.}\ \bibnamefont {Pollock}},\ }\bibfield  {title} {\bibinfo {title} {Temperature dependence of single crystal elastic constants in a coni-base alloy: A new methodology},\ }\href {https://doi.org/https://doi.org/10.1016/j.msea.2020.140507} {\bibfield  {journal} {\bibinfo  {journal} {Materials Science and Engineering: A}\ }\textbf {\bibinfo {volume} {803}},\ \bibinfo {pages} {140507} (\bibinfo {year} {2021})}\BibitemShut {NoStop}%
\bibitem [{\citenamefont {Li}\ \emph {et~al.}(2021)\citenamefont {Li}, \citenamefont {Said}, \citenamefont {Yan}, \citenamefont {Mandrus}, \citenamefont {Lee}, \citenamefont {Okamoto}, \citenamefont {Hal{\'a}sz},\ and\ \citenamefont {Miao}}]{li2021}%
  \BibitemOpen
  \bibfield  {author} {\bibinfo {author} {\bibfnamefont {H.}~\bibnamefont {Li}}, \bibinfo {author} {\bibfnamefont {A.}~\bibnamefont {Said}}, \bibinfo {author} {\bibfnamefont {J.}~\bibnamefont {Yan}}, \bibinfo {author} {\bibfnamefont {D.}~\bibnamefont {Mandrus}}, \bibinfo {author} {\bibfnamefont {H.}~\bibnamefont {Lee}}, \bibinfo {author} {\bibfnamefont {S.}~\bibnamefont {Okamoto}}, \bibinfo {author} {\bibfnamefont {G.~B.}\ \bibnamefont {Hal{\'a}sz}},\ and\ \bibinfo {author} {\bibfnamefont {H.}~\bibnamefont {Miao}},\ }\bibfield  {title} {\bibinfo {title} {{Divergence of Majorana-Phonon Scattering in Kitaev Quantum Spin Liquid}},\ }\href {https://arxiv.org/abs/2112.02015} {\bibfield  {journal} {\bibinfo  {journal} {arXiv preprint arXiv:2112.02015}\ } (\bibinfo {year} {2021})}\BibitemShut {NoStop}%
\end{thebibliography}%




\end{document}


\title{Supplemental Material for ``\textbf{\ourtitle}"}

\author{
Penghao Zhu$^{1}$, 
Archisman Panigrahi$^{2}$, 
Leonid Levitov$^{2}$, and 
Nandini Trivedi$^{1}$
}

\affiliation{$^1$Department of Physics, The Ohio State University, Columbus, OH 43210, USA\\
$^2$Department of Physics, Massachusetts Institute of Technology, 77 Massachusetts Avenue, MA 02139, USA}


\maketitle
\tableofcontents

\setcounter{section}{0}
\setcounter{figure}{0}
\setcounter{equation}{0}
\renewcommand{\thefigure}{S\arabic{figure}}
\renewcommand{\theequation}{S\arabic{equation}}
\renewcommand{\thesection}{S\arabic{section}}
\onecolumngrid
\section{Strain field in the Majorana-hopping model}
In the main text, we have demonstrated how strain affects the Majorana-hopping model in real space. To show that the strain field acts as an effective gauge field for low-energy Dirac modes, we first rewrite the Hamiltonian in momentum space and then expand both the momentum and strain field around a high-symmetry point, where there are low-energy Dirac modes.

Let us consider a honeycomb lattice with the following lattice vectors and reciprocal lattice vectors:
\begin{equation}
\label{eq:latticevector}
\begin{aligned}
&\text{Lattice vector:}\ \mathbf{a}_{1}=\frac{a}{2}\left(\sqrt{3},3\right)^{T}, \ \mathbf{a}_{2}=\frac{a}{2}\left(-\sqrt{3},3\right)^{T}
\\
&\text{Reciprocal lattice vector:}\ \mathbf{b}_{1}=\frac{2\pi}{a}\left(\frac{1}{\sqrt{3}},\frac{1}{3}\right)^{T}, \ \mathbf{a}_{2}=\frac{2\pi}{a}\left(-\frac{1}{\sqrt{3}},\frac{1}{3}\right)^{T}
\end{aligned}
\end{equation}
where $a$ is the separation between $A$ and $B$ sites, as shown in Fig.~1(b) in the main text. When there is no strain, we can Fourier transform the model in Eq. (3) by Fourier transforming the Majorana operators:
\begin{equation}
\label{eq:FTMajorana}
a^{\dag}_{\mathbf{k}}=\frac{1}{2\sqrt{N}}\sum_{i \in A \ \text{sites}}e^{i\mathbf{k}\cdot(\mathbf{R}_{i}+\boldsymbol{\delta}_3)}c_{i}, \ b^{\dag}_{\mathbf{k}}=\frac{1}{2\sqrt{N}}\sum_{i \in B \ \text{sites}}e^{i\mathbf{k}\cdot\mathbf{R}_{i}}c_{i},
\end{equation}
where $N$ is the number of total unit cells, $\mathbf{R}_{i}$ is the coordinate of the B site in the i-th unit cell, and $\boldsymbol{\delta}_{3}$ is the vector pointing from B site to A site within each unit cell [c.f. Fig.~1(b) in the main text]. Note that these complex fermionic operators satisfy the anticommutation relation:
\begin{equation}
\{\mathbf{a}^{\dag}_{\mathbf{k}},\mathbf{a}_{\mathbf{k}'}\}=\{\mathbf{b}^{\dag}_{\mathbf{k}},\mathbf{b}_{\mathbf{k}'}\}=\delta_{\mathbf{k}\mathbf{k}'}, \ \{\mathbf{a}^{\dag}_{\mathbf{k}},\ \mathbf{b}_{\mathbf{k}'}\}=0.
\end{equation}
Moreover, there is a particle-hole symmetry: 
$\mathbf{a}^{\dag}_{\mathbf{k}}=\mathbf{a}_{-\mathbf{k}}$ and $\mathbf{b}^{\dag}_{\mathbf{k}}=\mathbf{b}_{-\mathbf{k}}$.

With, Eq.~\eqref{eq:FTMajorana}, Eq.~(3) in the main text can be rewritten as
\begin{equation}
\label{eq:Hk}
H_{M}=\sum_{\mathbf{k}}\begin{pmatrix}
  \mathbf{a}^{\dag}_{\mathbf{k}}  & \mathbf{b}^{\dag}_{\mathbf{k}}  
\end{pmatrix}\begin{pmatrix}
M(\mathbf{k})& iJ_{0}\sum_{\alpha=1}^{3}e^{i\mathbf{k}\cdot\boldsymbol{\delta}_{\alpha}}
 \\
-i J_{0}\sum_{\alpha=1}^{3}e^{-i\mathbf{k}\cdot\boldsymbol{\delta}_{\alpha}} &  -M(\mathbf{k})
\end{pmatrix}\begin{pmatrix}
    \mathbf{a}_{\mathbf{k}}
    \\
    \mathbf{b}_{\mathbf{k}}
\end{pmatrix}=\sum_{\mathbf{k}}E_{\mathbf{k}}(\xi^{\dag}_{+,\mathbf{k}}\xi_{+,\mathbf{k}}-\xi^{\dag}_{-,\mathbf{k}}\xi_{-,\mathbf{k}})
\end{equation}
with $M(\mathbf{k})= 2\lambda(\sin \mathbf{k}\cdot\mathbf{a}_1-\sin\mathbf{k}\cdot \mathbf{a}_2-\sin\mathbf{k}\cdot(\mathbf{a}_{1}-\mathbf{a}_2))$, $E_{\mathbf{k}}\geqslant 0$, and 
\begin{equation}
\boldsymbol{\delta}_1=\frac{a}{2}(\sqrt{3},-1)^{T}, \ \boldsymbol{\delta}_2=\frac{a}{2}(-\sqrt{3},-1)^{T}, \
\boldsymbol{\delta}_3=a(0,1)^{T}.
\end{equation}
Due to the particle-hole symmetry, there is a redundancy such that $\xi^{\dag}_{+,\mathbf{k}}=\xi_{-,-\mathbf{k}}$ and $E_{\mathbf{k}}=E_{-\mathbf{k}}$. As a result, 
\begin{equation}
\label{eq:Hk1}
H_{M}=\sum_{\mathbf{k}}E_{\mathbf{k}}(\xi^{\dag}_{+,\mathbf{k}}\xi_{+,\mathbf{k}}-\xi^{\dag}_{-,-\mathbf{k}}\xi_{-,-\mathbf{k}})=\sum_{\mathbf{k}}E_{\mathbf{k}}(2\xi^{\dag}_{+,\mathbf{k}}\xi_{+,\mathbf{k}}-1),
\end{equation}
which has a ground state with all states with negative energy occupied. The ground state energy is given by $-\sum_{\mathbf{k}}E_{\mathbf{k}}$, and the energy of a single-particle excitation is given by $\epsilon_{\mathbf{k}}=2E_\mathbf{k}$. The factor $2$ in the excitation energy motivates us to study the following Hamiltonian in first quantization language:
\begin{equation}
h_{\mathbf{k}}=2\begin{pmatrix}
M(\mathbf{k})& iJ_{0}\sum_{\alpha=1}^{3}e^{i\mathbf{k}\cdot\boldsymbol{\delta}_{\alpha}}
 \\
-i J_{0}\sum_{\alpha=1}^{3}e^{-i\mathbf{k}\cdot\boldsymbol{\delta}_{\alpha}} &  -M(\mathbf{k})
\end{pmatrix},
\end{equation}
of which the $\mathbf{k}\cdot \mathbf{p}$ expansion around high symmetry momenta $K=-K'=-2\pi/a(1/(3\sqrt{3}),1/3)$ is exactly Eq.~(4) in the main text.

Let us now consider small a strain field $\epsilon$ that modifies the exchange interaction $J_{0}$ as follows:
\begin{equation}
J_{\alpha} = J_{0}\exp\left[-\beta\left(\frac{|(I + \epsilon)\boldsymbol{\delta}_{\alpha}|}{a} - 1\right)\right]\approx J_{0}\left[1-\beta\left(\frac{|(I + \epsilon)\boldsymbol{\delta}_{\alpha}|}{a} - 1\right)\right].
\end{equation}
Note that the strain field $\epsilon$ in the discrete honeycomb lattice is defined through the discrete derivative of the displacement $\mathbf{s}$ -- The strain field at the i-th unit cell with coordinate $\mathbf{R}_{i}$ can be defined through
\begin{equation}
  [(\boldsymbol{\delta}_{\alpha}\cdot \nabla)s_{j}](\mathbf{R}_{i})\equiv[(\boldsymbol{\delta}_{\alpha}^{t}\epsilon)_{j}](\mathbf{R}_{i})=\frac{s_{j}(\mathbf{R}_{i}+\boldsymbol{\delta}_{\alpha})-s_{j}(\mathbf{R}_{i})}{a}.
\end{equation}
Around the high symmetry momenta $K$ and $K'$, up to the linear order in $\epsilon$, this modification leads to two effects to the Majorana hopping model: Firstly, a correction to Dirac mass through the next nearest neighbor hopping strength $\lambda\sim h^3/(J_{\alpha_{jl}}J_{\alpha_{lk}})$. Secondly, an effective gauge field minimally coupled to the momentum. To see these effects, let us focus on the $K$ point and expand the Hamiltonian up to linear order in both momentum and strain field $\epsilon$. It is important to note that, in general, strain breaks translational symmetry. However, we can still use momentum as a good quantum number under the assumption that the strain field varies slowly in space. In this regime, the system can be viewed as consisting of many large patches with approximately uniform strain, within which translational symmetry is effectively preserved. The linear size of these patches, denoted by $L_{\text{patch}}$, sets a lower bound on the momentum resolution, such that the minimal momentum we can meaningfully consider is $k_{\text{min}} \sim 1 / L_{\text{patch}}$.

We start by looking at the diagonal term $M(\mathbf{k})$ and note that
\begin{equation}
\frac{|(I + \epsilon)\boldsymbol{\delta}_{\alpha}|}{a} - 1=\frac{\sqrt{\boldsymbol{\delta}_{\alpha}^{T}(I+\epsilon)^{T}(I+\epsilon)\boldsymbol{\delta}_\alpha}}{a}-1\approx\sqrt{1+\boldsymbol{\delta}_{\alpha}^{T}(\epsilon^{T}+\epsilon)\boldsymbol{\delta}_\alpha/a^2}-1\approx\frac{1}{2}\frac{\boldsymbol{\delta}_{\alpha}^{T}(\epsilon^{T}+\epsilon)\boldsymbol{\delta}_\alpha}{a^2}.
\end{equation}
Then,
\begin{equation}
\begin{aligned}
M(K+\mathbf{k}) &=2\lambda_{0}\Bigg[\left(1+\frac{\beta}{2}\frac{\boldsymbol{\delta}_{2}^{T}(\epsilon^{T}+\epsilon)(\mathbf{R}_{i})\boldsymbol{\delta}_2}{a^2}+\frac{\beta}{2}\frac{\boldsymbol{\delta}_{3}^{T}(\epsilon^{T}+\epsilon)(\mathbf{R}_{i})\boldsymbol{\delta}_3}{a^2}\right)\left(\frac{\sqrt{3}}{2}-\frac{1}{2}\mathbf{k}\cdot\mathbf{a}_1\right)
\\
&+\left(1+\frac{\beta}{2}\frac{\boldsymbol{\delta}_{1}^{T}(\epsilon^{T}+\epsilon)(\mathbf{R}_{i}+\mathbf{a}_2-\mathbf{a}_1)\boldsymbol{\delta}_1}{a^2}+\frac{\beta}{2}\frac{\boldsymbol{\delta}_{3}^{T}(\epsilon^{T}+\epsilon)(\mathbf{R}_{i}+\mathbf{a}_2-\mathbf{a}_1)\boldsymbol{\delta}_3}{a^2}\right)\left(\frac{\sqrt{3}}{2}+\frac{1}{2}\mathbf{k}\cdot\mathbf{a}_2\right)
\\
&+\left(1+\frac{\beta}{2}\frac{\boldsymbol{\delta}_{1}^{T}(\epsilon^{T}+\epsilon)(\mathbf{R}_{i}+\mathbf{a}_2)\boldsymbol{\delta}_1}{a^2}+\frac{\beta}{2}\frac{\boldsymbol{\delta}_{2}^{T}(\epsilon^{T}+\epsilon)(\mathbf{R}_{i}+\mathbf{a}_2)\boldsymbol{\delta}_2}{a^2}\right)\left(\frac{\sqrt{3}}{2}+\frac{1}{2}\mathbf{k}\cdot(\mathbf{a}_1-\mathbf{a}_2)\right)\Bigg]+O(\epsilon^2)+O(|\mathbf{k}|^2).
\end{aligned}
\end{equation}
Remember the strain field varies slowly spatially, and thus we can safely drop the position dependence and write the above equation as

\begin{equation}
\begin{aligned}
M(K+\mathbf{k}) \approx3\sqrt{3}\lambda_{0}(1+3\epsilon_{xx}+3\epsilon_{yy})\equiv m',
\end{aligned}
\end{equation}
which reveals a linear correction to the Dirac mass induced by the strain field.

Next, let us expand the off-diagonal term
\begin{equation}
\begin{aligned}
i \sum_{\alpha=1}^{3}J_{\alpha}e^{i\mathbf{k}\cdot\boldsymbol{\delta}_{\alpha}}&=J_0\Bigg[\left(1-\frac{\beta}{2}\frac{\boldsymbol{\delta}_{1}^{T}(\epsilon^{T}+\epsilon)(\mathbf{R}_{i}+\mathbf{a}_2)\boldsymbol{\delta}_1}{a^2}\right)(1+i\mathbf{k}\cdot\boldsymbol{\delta_{1}})
\\
&+e^{i2\pi/3}\left(1-\frac{\beta}{2}\frac{\boldsymbol{\delta}_{2}^{T}(\epsilon^{T}+\epsilon)(\mathbf{R}_{i}+\mathbf{a}_2)\boldsymbol{\delta}_2}{a^2}\right)(1+i\mathbf{k}\cdot\boldsymbol{\delta_{2}})
\\
&+e^{-i2\pi/3}\left(1-\frac{\beta}{2}\frac{\boldsymbol{\delta}_{3}^{T}(\epsilon^{T}+\epsilon)(\mathbf{R}_{i}+\mathbf{a}_2)\boldsymbol{\delta}_3}{a^2}\right)(1+i\mathbf{k}\cdot\boldsymbol{\delta_{3}})+O(\epsilon^2)+O(|\mathbf{k}|^2)
\\
&\approx-\frac{3J_{0}a}{2}ie^{-i2\pi/3}\left[k_{x}-i k_{y}+\frac{\beta}{2a}(\epsilon_{yy}-\epsilon_{xx})-i\frac{\beta}{2a}(\epsilon_{xy}+\epsilon_{yx})\right].
\end{aligned}
\end{equation}
Comparing the above equation to the minimal coupling between electron and $U(1)$ gauge field, $\mathbf{k}\rightarrow \mathbf{k}+\frac{e}{\hbar}\mathbf{A}$ and remembering that the strain field is symmetric, we can extract the effective gauge
\begin{equation}
\mathbf{A}_{\text{eff}}=-\frac{\beta}{2 a}\frac{\hbar}{e}(\epsilon_{xx}-\epsilon_{yy}, -2\epsilon_{xy}),
\end{equation}
which is exactly Eq.~(6) in the main text. Up to a unitary gauge transformation corresponding to basis change, the effective low energy Hamiltonain around K in a strained system is given by
\begin{equation}
    h_{\text{Dirac}}^{\text{strained}}=3J_{0}a\left(\mathbf{k}+\frac{e}{\hbar}\mathbf{A}_{\text{eff}}\right)\cdot\boldsymbol{\sigma}+m'\sigma_z,
\end{equation}
where $\boldsymbol{\sigma}=(\sigma_x,\sigma_y)$, and $h_{\text{Dirac}}^{\text{strained}}$ supports pseudo-Landau levels.

\section{More numerical data about LDOS in the Majorana metal phase}
\begin{figure}[h]
\centering
\includegraphics[width=0.8\columnwidth]{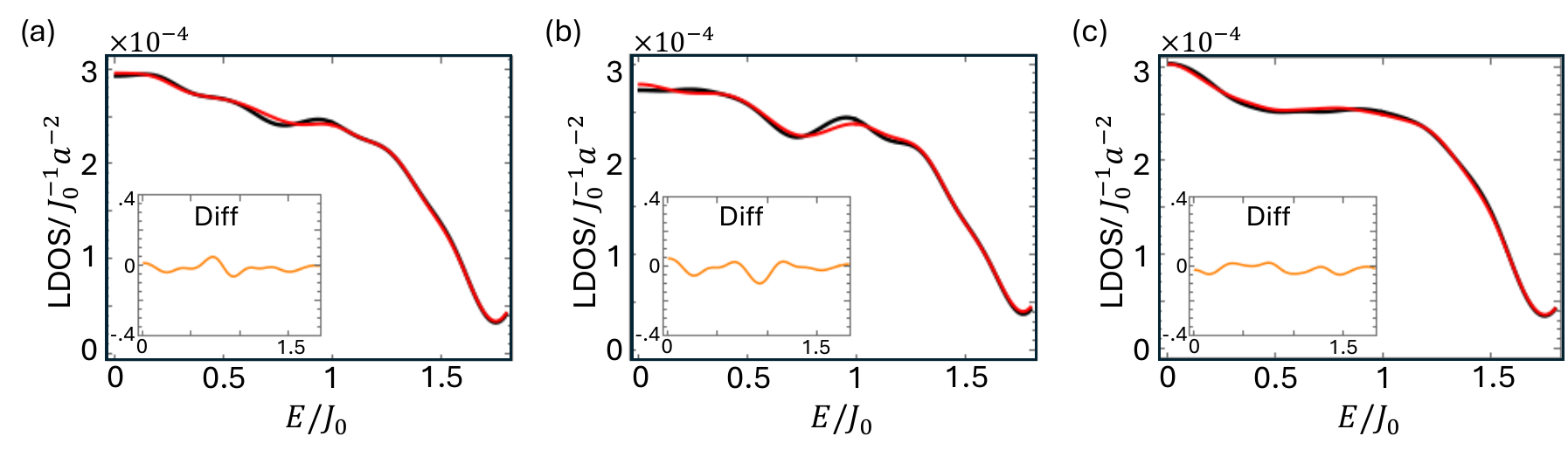}
\caption{Local density of states (LDOS) for region $\mathcal{R}$, defined as the central hexagon, in the Majorana metal phases calculated for three different sample averages. Red lines correspond to strained systems with $\beta \mathcal{S} = 0.4$, while black lines represent unstrained systems ($\beta \mathcal{S} = 0$). Insets show the LDOS difference between strained and unstrained systems. Each plot is produced by averaging over 50 samples.}
\label{figS1}
\end{figure}

In Fig.~\ref{figS1}, we present the local density of states (LDOS) obtained by averaging over three different sets of disordered samples. The results show that the difference between the strained and unstrained LDOS exhibits oscillations that depend on sample averaging. These oscillations have relatively small amplitudes compared to those observed in the KSL and CSL phases, and the spacing between peaks does not follow any discernible pattern. Thus, we conclude there is no pseudo-Landau levels in the Majorana metal phase.

\section{Dispersion of low energy modes in the Majorana metal phase}

We determine the dispersion of low energy modes around the $K$ point in the Majorana metal phase by studying the peak of the spectral function. Following Ref.~\cite{zhu2025}, we calculate the spectral function for the Majorana-hopping model with sign disorder using 
\begin{equation}
\label{eq:specfunck}
 \begin{split}
A(\mathbf{k},\omega)&=\sum_{n\alpha}|c^{n}_{\alpha\mathbf{k}}|^2\delta(\omega - E_n),
    \end{split}
\end{equation}
where $c^{n}_{\alpha\mathbf{k}}\equiv\langle\phi_n|\alpha\mathbf{k}\rangle$ is the overlap between the n-th eigenstates, $\ket{\phi_{n}}$, of the disordered Majorana-hopping model with energy and the eigenstates, $\ket{\alpha\mathbf{k}}$. of the clean system. This calculation is performed across different disordered flux patterns, and the resulting spectral functions are averaged over samples.

Having computed the sample-averaged spectral function $\langle A(\mathbf{k}, \omega) \rangle_{\text{sample}}$, we identify, for each momentum $\mathbf{k}$ near the $K$ point, the location of its peak within the energy window below the first gap, i.e., $\omega \in [0, 2]$. The position of this peak provides the energy of the mode corresponding to momentum $\mathbf{k}$, allowing us to extract the effective dispersion of low-energy modes around the $K$ point, as shown in the main text.

\section{Landau diamagnetism of Dirac modes}
\textit{Goal}: In this section, we will derive the expression of magnetic susceptibility of a non-interacting many-body Fermionic system with charged quasiparticles having a massive Dirac dispersion, under the condition of net charge neutrality.

\begin{figure}[h]
    \centering
    \includegraphics[width=0.8\linewidth]{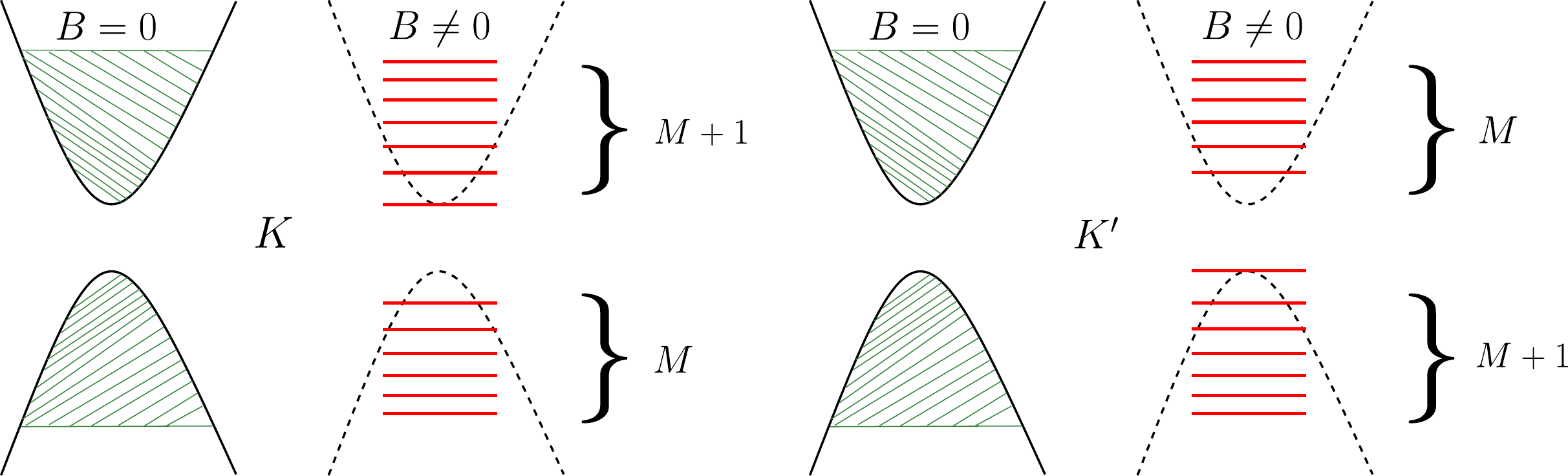}
    \caption{As a finite magnetic field $B$ is turned on, we consider $(2M + 1)$ low-lying Landau levels in each of $K$ and $K'$ valleys. We denote the free-energy of the electrons occupying these states as $\Omega(B)$. Next, we consider the states at zero magnetic field (marked with green), which combine to form these $2M+1$ Landau levels at the finite magnetic field. We denote the free-energy of these states as $\Omega(0)$. Finally, we will calculate the difference $\Omega(B) - \Omega(0)$ and take the $M \rightarrow \infty$ limit to isolate the susceptibility, which remains finite. If we take the limit $M\rightarrow \infty$ without calculating the change in free energy, both the quantities $\Omega(B)$ and $\Omega(0)$ individually diverge (but their difference remains finite).}
    \label{fig:landa-levels}
\end{figure}

The dispersion of massive Dirac modes (Eq. (4) in the main text, with $v_{\rm eff}=3 J_0 a/\hbar$) is \begin{equation}
    \epsilon(\vec k) = \pm \sqrt{m^2 + v_{\rm eff}^2 k^2 \hbar^2}
\end{equation} and the corresponding density of states is,
\begin{equation}
    g(\epsilon) = \frac{\epsilon}{2\pi v_{\rm eff}^2 \hbar^2}.
\end{equation}
In the $K$ valley, the energies of the Landau levels are,
$\epsilon_{0,K} = m$, and $\epsilon_{n,\pm,K} =\pm \sqrt{m^2 +2 v_{\rm eff}^2 \hbar eB n}$ (where $n \in \mathbb{Z}^+$). 
In the $K'$ valley, the energies of the Landau levels are,
$\epsilon_{0,K'} = -m$, and $\epsilon_{n,\pm,K'} =\pm \sqrt{m^2 +2 v_{\rm eff}^2 \hbar eB n}$. Here $e$ denotes the charge of a proton.
Each Landau level has degeneracy $\Phi/\Phi_0 = \frac{B A e}{2\pi \hbar}$.

In order to calculate susceptibility, we need to first compute the grand potential (or Landau free energy),
\begin{equation}\label{eq:landau-free-energy}
\Omega = - k_B T \sum_n g_n \log\left[1 + e^{\beta (\mu - \epsilon_n)}\right],
\end{equation}
where $g_n$ is the degeneracy of the energy level $\epsilon_n$.
The sum in Eq.\eqref{eq:landau-free-energy} diverges due to all the filled, negative energy states. But it also diverges when there is no magnetic field. We are not interested in the divergent sum itself, but in the susceptibility, 
\begin{equation}
    \chi = -\frac{1}{A}\left. \frac{\partial^2 \Omega(B)}{\partial B^2}\right|_{B=0},
\end{equation} which should be finite.

We will regularize the sum, by calculating the energy change $\Omega(B) - \Omega(0)$ due to the contribution of the zeroeth Landau level and $M$ Landau levels below and $M$ Landau levels above the zero energy, for each valley. We will isolate the $B$-dependent contribution (which is finite, and $M$-independent when $M$ is large). Finally we will take the $M \rightarrow \infty$ limit, and differentiate twice to obtain the susceptibility.

In this section, we will do this calculation in the undoped limit, i.e. $\mu = 0$.

Combining the energy levels in the $K$ and $K'$ valleys,
\begin{equation}
\begin{aligned}
    \frac{\Omega(B)}{-k_B T} = \frac{B e}{2\pi \hbar} \Bigg[\log\left(1 + e^{\beta m}\right) + \log\left(1 + e^{-\beta m}\right)+ 2 \sum_{n=1}^M \log\left(1 + e^{\beta \sqrt{m^2 + 2 v_{\rm eff}^2 \hbar eB n}}\right) + \log\left(1 + e^{-\beta \sqrt{m^2 + 2 v_{\rm eff}^2 \hbar eB n}}\right) \Bigg]
\end{aligned}
\end{equation}
For $\Omega(B=0)$, we will include those energy-levels around zero energy, which, in each valley, contain same number of states as these $(2M+1)$ Landau levels.
At zero magnetic field, the grand potential of these energy levels is given by (combining contribution from both the valleys),
\begin{equation}
\begin{aligned}
    \Omega(0) &= -2 k_B T \times  \Bigg[\int_{m}^{E_M} d\epsilon A g(\epsilon) \log(1 + e^{-\beta \epsilon} ) +\int_{-E_M}^{-m} d\epsilon A g(\epsilon)  \log(1 + e^{-\beta \epsilon})\Bigg]\\
    &=-2 k_B T \Bigg[\int_{m}^{E_M} d\epsilon A g(\epsilon) \left[\log(1 + e^{\beta \epsilon} )+  \log(1 + e^{-\beta \epsilon})\right] \Bigg]
\end{aligned}
\end{equation}
where $E_M$ is such that the energy levels from $-E_M$ to $-m$ and $m$ to $E_M$ contain same number of states as the $(2M+1)$ Landau levels, i.e.,
\begin{equation}
    \int_m^{E_M} d\epsilon A g(\epsilon) + \int_{-E_M}^{-m} d\epsilon A g(\epsilon) = (2M+1) \frac{BAe}{2\pi \hbar},
\end{equation}
which gives, 
\begin{equation}
    E_M = \sqrt{m^2 + (2M+1)v_{\rm eff}^2 \hbar eB}.
\end{equation}
Substituting this in the expression of $\Omega(0)$, and changing the integration variable from $\epsilon$ to $n = \frac{\epsilon^2 - m^2}{2 v_{\rm eff}^2 \hbar B e}$, we obtain,
\begin{equation}\label{eq:omega0-in-terms-of-Landau-level-index}
\begin{aligned}
    \Omega(0) = (-2 k_B T) A \frac{B e}{2\pi \hbar} \int_0^{M+\frac{1}{2}} dn  \Big[\log\left(1 + e^{\beta \sqrt{m^2 + 2 v_{\rm eff}^2 \hbar eB n}}\right) + \log\left(1 + e^{-\beta \sqrt{m^2 + 2 v_{\rm eff}^2 \hbar eB n}}\right) \Big].
\end{aligned}
\end{equation}
Note that the actual $\Omega(0)$ itself has no magnetic field dependence, and it contains all energy levels from $-\infty$ to $\infty$ (indeed, the $B$-dependence in Eq.\eqref{eq:omega0-in-terms-of-Landau-level-index} drops out after taking $M\rightarrow \infty$ limit). Here, to regularize $\Omega$, we calculated the contribution to free energy solely due to those energy levels which would eventually form the $(2M+1)$ Landau levels closest to charge neutrality. Finally we will take the $M \rightarrow \infty$ limit when we will calculate the change in grand potential due to the magnetic field. Let us define,
\begin{equation}
\begin{aligned}
    h(n) = \log\left(1 + e^{\beta \sqrt{m^2 + 2 v_{\rm eff}^2 \hbar eB n}}\right) + \log\left(1 + e^{-\beta \sqrt{m^2 + 2 v_{\rm eff}^2 \hbar eB n}}\right).
\end{aligned}
\end{equation}
Therefore,
\begin{equation}\label{eq:diff-free-energy}
\begin{aligned}
    \frac{\Omega(B) - \Omega(0)}{-A k_B T}
    = \frac{Be}{2\pi \hbar} \left[h(0) + 2 \sum_{n=1}^M h(n) - 2 \int_0^{M+\frac{1}{2}} h(n) dn\right].
\end{aligned}
\end{equation}
\underline{Until here}, the expressions were \textit{exact}. To calculate the susceptibility, we need to expand this quantity upto quadratic order in $B$, for which, we will utilize the Euler-Maclaurin formula to calculate the difference between the sum and the integral,
\begin{equation}\label{eq:Euler-Maclaurin-series}
    \sum_{n=1}^M h(n) = \int_1^M h(n) dn + \frac{h(1) + h(M)}{2} + \frac{h'(M)-h'(1)}{12} + \cdots.
\end{equation}
The higher order terms in the series will give rise to cubic and higher powers in $B$ (each additional derivative w.r.t. $n$ gives rise to an additional factor of $B$). For our purpose, it is enough to truncate the series at this stage.
Substituting Eq.\eqref{eq:Euler-Maclaurin-series} in Eq.\eqref{eq:diff-free-energy}, we obtain,
\begin{equation}
\begin{aligned}
    \frac{\Omega(B) - \Omega(0)}{-A k_B T} = &\frac{B e}{2\pi \hbar} \Bigg[h(0) + h(1) - 2 \int_0^{1} h(n) dn - \frac{h'(1)}{6}\Bigg]
    + \frac{B e}{2\pi \hbar} \underbrace{\Bigg[h(M) - 2 \int_M^{M+\frac{1}{2}} h(n) dn + \frac{h'(M)}{6}\Bigg]}_{\text{we'll show it approaches 0 for large $M$}}\\
    &\qquad+ O(B^3).
\end{aligned}
\end{equation}
We note that
\begin{equation}
\begin{aligned}
        h(M) - 2 \int_M^{M+\frac{1}{2}} h(n) dn &= -\frac{1}{4}h'(M) -\frac{1}{24}h''(M) + \cdots\\
        &\quad \rightarrow 0 \text{ as $M \rightarrow \infty$,}
\end{aligned}
\end{equation}
because
\begin{equation}
        h'(M) = \frac{\beta v_{\rm eff}^2 \hbar eB}{\sqrt{m^2 + v_{\rm eff}^2 \hbar eB M}} \tanh\left(\frac{\beta}{2}\sqrt{m^2 + 2 v_{\rm eff}^2 \hbar eB M}\right) \rightarrow 0
\end{equation}
as $M \rightarrow \infty$, for any finite $B$ (and it is identically zero for $B=0$). We will take the large $M$ limit and drop these terms. Physically, the relation says that, the contribution of the high-energy states to the change in grand potential (hence, to the physical observables) is negligible. All the contribution comes from the low-lying states.
Therefore, taking contributions from all the Landau levels ($M\rightarrow \infty$),
\begin{equation}
\begin{aligned}
    \frac{\Omega(B) - \Omega(0)}{-A k_B T} = &\frac{B e}{2\pi \hbar} \Bigg[h(0) + h(1) - 2 \int_0^{1} h(n) dn - \frac{h'(1)}{6}\Bigg] \\
    & \qquad + O(B^3),
\end{aligned}
\end{equation}
Note that we obtained this expression by adding contributions from all Landau levels (the high energy-levels have little contribution), and not just from one or two lowest Landau levels.
To get the contribution upto quadratic order in $B$, we linearize $h(n) \approx h(0) + n h'(0) = h(0) + n \frac{\beta v_{\rm eff}^2 \hbar eB}{m} \tanh\left(\frac{\beta m}{2}\right)$, and obtain,
\begin{equation}
\begin{aligned}
    \frac{\Omega(B) - \Omega(0)}{-A k_B T} &= \frac{B e}{2\pi \hbar} \left(- \frac{h'(0)}{6}\right) + O(B^3)\\
    &=-\frac{v_{\rm eff}^2 e^2}{12\pi k_B T} \frac{\tanh\left(\frac{\beta m}{2}\right)}{m} B^2 + O(B^3).
\end{aligned}
\end{equation}
Using the relation $\chi = -\frac{1}{A} \left.\frac{\partial^2 \Omega(B)}{\partial B^2} \right|_{B=0}$, we obtain,
\begin{equation}
\label{eq:suscsl}
       \boxed{\chi = - \frac{v_{\rm eff}^2 e^2}{6\pi} \frac{\tanh\left(\frac{m}{2 k_B T}\right)}{m} }.
\end{equation}
It is interesting to note that $\hbar$ drops out of the final result.

\section{Response to strain of higher energy modes}

Here, we demonstrate that the mode around the $M=(2\pi/3,0)$ point has a paramagnetic contribution to the strain susceptibility.  We now expand the Hamiltonian in Eq.~\eqref{eq:Hk} to the lowest order in $\epsilon$ and deviation from $M$, $\mathbf{q}$:
\begin{equation}
\label{eq:expand}
h=\begin{pmatrix}
2\sqrt{3}\lambda_{0}q_{y} & e^{i\pi/3}t_{0}[1+2iq_x-\beta(3\epsilon_{yy}/2-\epsilon_{xx}/2)]
\\
e^{-i\pi/3}t_{0}[1-2iq_x-\beta(3\epsilon_{yy}/2-\epsilon_{xx}/2)] & -2\sqrt{3}\lambda_{0}q_{y}, 
\end{pmatrix}
\end{equation}
It is straight forward to see that the strain field here does not behave like an effective gauge field as for the modes around $K$. This Hamiltonian has energy spectrum 
\begin{equation}
E=\pm \sqrt{12\lambda_{0}^2q_{y}^2+t_{0}^2[1-\beta(3\epsilon_{yy}/2-\epsilon_{xx}/2)]^2+4t_{0}^2q_x^2}.
\end{equation}

Given that $\epsilon_{xx}=2\mathcal{S}R_y/L$ and $u_{yy}=-2\mathcal{S}R_{y}/L$, 

\begin{equation}
\begin{aligned}
E&=\pm \sqrt{12\lambda_{0}^2q_{y}^2+t_{0}^2[1+4\beta\mathcal{S}R_y/L]^2+4t_{0}^2q_x^2}
\\
&=\pm \sqrt{12\lambda_{0}^2q_{y}^2+t_{0}^2[1+8\beta\mathcal{S}R_y/L+16(\beta\mathcal{S}R_y/L)^2]+4t_{0}^2q_x^2}
\\
&\approx \pm t_{0}[1+(6\lambda_{0}^2q_{y}^2+2t_{0}^2q_x^2+4\beta\mathcal{S}R_y/L+8(\beta\mathcal{S}R_y/L)^2)/t_{0}^2].
\end{aligned}
\end{equation}
The local region we focus on has sites sitting symmetrically with respect to $y=0$. As a result the term linear in $\mathcal{S}$ should be canceled out, and the strain field decreases the the energy of states with negative energy. Therefore, the strain will lower the ground state energy with $\chi_{\text{strain}}=-\partial^2 E_G/\partial \mathcal{S}^2>0$, which is paramagnetic.

\section{Details of effective magnetization and strain susceptibility}

\begin{figure}[h]
\centering
\includegraphics[width=0.8\columnwidth]{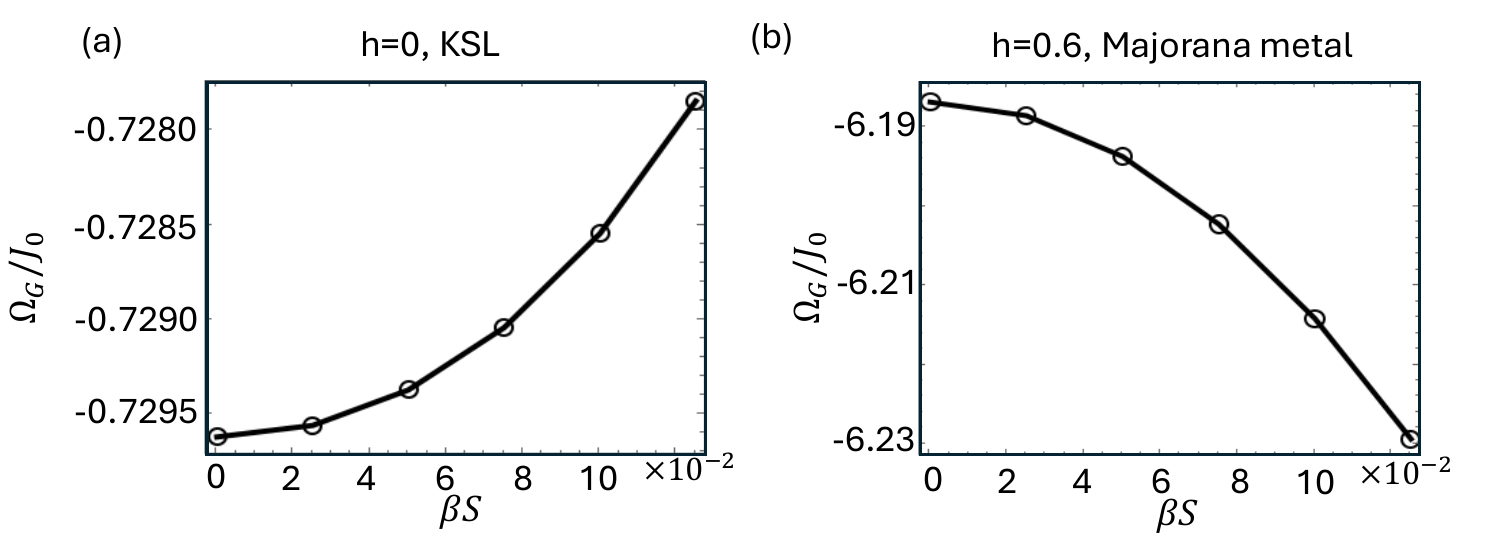}
\caption{Grand potentials calculated for the spin model with \( 4 \times 2 \) unit cells and open boundary conditions under (a) \( h = 0 \) and (b) \( h = 0.6 J_0 \). To avoid large exponentials in \( e^{-\epsilon_n / (k_B T)} \), the energy spectrum is uniformly shifted by \( 10 J_0 \) in the calculations.}
\label{figS2}
\end{figure}

The ED calculations of the spin model at \( h = 0 \) and \( h = 0.6 J_0 \), performed at temperature \( T = 0.05 J_0 \), yield grand potentials shown in Fig.~\ref{figS2}. In both cases, the grand potential exhibits a quadratic dependence on the strain field strength \( \mathcal{S} \), indicating that the effective zero-field magnetization vanishes. This allows us to focus on the strain susceptibility as the key quantity characterizing the response to strain in these phases.

We note that this quadratic dependence is well captured by the Majorana-hopping model in the KSL and CSL phases, as discussed in the main text. However, in the Majorana metal phase, the behavior differs: the sample-averaged grand potential does not consistently exhibit a quadratic dependence on \( \mathcal{S} \). Instead, its behavior is highly sensitive to the specific sample realization. This suggests that while the disordered Majorana-hopping model reproduces the gapless nature of the low-energy spectrum in the Majorana-metal phase, it does not reliably capture the effective magnetization or strain susceptibility. Accurately determining these quantities requires detailed knowledge of the ground-state flux configuration in the full spin model.




\bibliography{refs}